\newif\ifnotes
\notestrue %

\newif\ifpnas
\pnasfalse

\documentclass[letterpaper]{article}
\usepackage[utf8]{inputenc}
\usepackage[at]{easylist}
\usepackage{graphicx}
\usepackage{caption}
\usepackage{subcaption}
\usepackage{float}
\usepackage{booktabs}
\usepackage[numbers]{natbib} %
\usepackage[margin=1in]{geometry}
\usepackage[hidelinks]{hyperref}
\usepackage{xspace}
\usepackage[table]{xcolor}
\usepackage{dsfont}
\usepackage{pifont}
\usepackage{wrapfig}
\usepackage{authblk}
\usepackage{multicol}

\usepackage{enumitem}
\setdescription{leftmargin=0pt}

\usepackage{amsfonts,amsmath,amssymb,amsthm}
\usepackage{booktabs} %
\usepackage[normalem]{ulem} %

\newcommand{\etal}{\textit{et al.}\xspace}

\newcommand{\plan}{\mathcal{P}}
\newcommand{\mmd}{\mathsf{MMD}}
\newcommand{\dist}{D}

\newcommand{\pop}{\mathsf{pop}}
\newcommand{\bpop}{\mathsf{pop}^\mathsf{Black}}
\newcommand{\bvap}{\mathsf{BVAP}}
\newcommand{\vap}{\mathsf{VAP}}
\newcommand{\bvapmargin}{B}%

\newcommand{\numDist}{k}
\newcommand{\idealPop}{\overline{p}}
\newcommand{\dev}{\mathsf{dev}}

\newcommand{\ddp}{\mathsf{demo}}
\newcommand{\data}{\mathsf{data}}
\newcommand{\sfo}{\mathsf{swap}}
\newcommand{\error}{\mathsf{err}_{\mathsf{das}}}

\newcommand{\normal}{\mathrm{N}}

\newcommand{\recom}{\ensuremath{\mathsf{ReCom}}\xspace}

\newcommand{\vanilla}{base\xspace}
\newcommand{\bursts}{short bursts\xspace}

\newcommand{\tol}{\tau}
\newcommand{\offset}{\Delta}

\usepackage{xcolor}
\ifnotes
    \newcommand{\mynote}[3]{\marginpar{{\tiny \parbox{0.8in}{\color{#2} \sf {#1}: {#3}}}}}
    \newcommand{\inlinenote}[3]{{\bfseries \color{#2} {#1}: #3}}
    
\else
    \newcommand{\mynote}[3]{}
    \newcommand{\inlinenote}[3]{}
    
\fi

\newcommand{\hide}[1]{\mynote{Some text hidden here.}{red}{}}

\newcommand{\nt}[1]{{\color{teal}{#1}}}
\newcommand{\ot}[1]{{\color{lightgray}{\sout{#1}}}}

\renewcommand{\ot}[1]{}%
\renewcommand{\nt}[1]{#1} %

\title{Understanding and Mitigating the Impacts of Differentially Private Census Data on State Level Redistricting}
\author[1]{Christian Cianfarani}
\author[1]{Aloni Cohen}
\affil[1]{Department of Computer Science, University of Chicago}
\date{\texttt{\{crc,aloni\}@uchicago.edu}}

\begin{document}

\maketitle

\begin{abstract}
Data from the Decennial Census is published only after applying a disclosure avoidance system (DAS). Data users were shaken by the adoption of differential privacy in the 2020 DAS, a radical departure from past methods. 
The goal of this paper is to better understand how the perturbations from the 2020 DAS combine with sharp legal thresholds to impact redistricting.
We consider two redistricting settings in which a data user might be concerned about the impacts of privacy preserving noise: drawing equal population districts and litigating voting rights cases.
What discrepancies arise if the user does nothing to account for disclosure avoidance? 
How can the discrepancies be understood and accounted for?
We study these questions by comparing the official 2010 Redistricting Data to the 2010 Demonstration Data---created using the 2020 DAS---in an analysis of millions of algorithmically generated state legislative redistricting plans.  
We find that thresholding can amplify the impact of the noise from disclosure avoidance. Large discrepancies do occur, but in ways that are well-captured by simple models and appear to be possible to account for. We demonstrate the utility of these models by proposing an approach to mitigate discrepancies when balancing district populations.
At least for state legislatures, Alabama's claim that differential privacy ``inhibits a State's right to draw fair lines'' lacks support.
\end{abstract}

\ifpnas
\dropcap{I}n
\else
\section{Introduction}
\label{sec:intro}
In
\fi
 2021, the state of Alabama sued the US Department of Commerce. Alabama sought to enjoin the use of a new disclosure avoidance system for the forthcoming release of 2020 decennial census data.
\begin{quote}
``Forcing the State to redistrict with intentionally flawed data will impede Alabama's ability to draw representative districts with near-equal populations, which is what the Constitution and one-person, one-vote jurisprudence require. This will also impede Alabama's ability to draw districts to protect minority voting rights as required by the Voting Rights Act.'' (\emph{Alabama v.\ Department of Commerce}, 2021)
\end{quote}
\parshape=0
Concerns persisted even after the data were published. In early 2022, one of us was contacted by lawyers at the ACLU in connection with racial gerrymandering cases in Georgia and Arkansas. %
They wondered whether the new disclosure avoidance system could affect the reported number of majority-Black districts in a demonstration redistricting plan.\footnote{Ultimately, this issue was not raised by defendants. No expert testimony was needed.}

Every ten years, the US Census Bureau conducts a decennial census and produces the \emph{Redistricting Data Summary Files}---the basis for all redistricting for the following decade.
These Redistricting Data are created by first aggregating data from the \emph{Census Edited File} (CEF)---the confidential dataset that reflects the results of the completed census---and then applying the \emph{Disclosure Avoidance System} (DAS).
Through 2010, the DAS comprised traditional disclosure limitation techniques,
primarily swapping \citep{censushistory}. These techniques are now understood as lacking formal guarantees against re-identification and individualized data disclosures \citep{garfinkel2018understanding,dick2023confidence}.

In contrast, the 2020 DAS is based on differential privacy, a modern framework for formally quantifying and limiting such disclosure risks \citep{dwork2006calibrating, Abowd20222020}.
The 2020 DAS synthesizes and aggregates a new set of microdata census responses that have no one-to-one correspondence with the actual microdata in the CEF but closely approximate many of its statistics.
While previous DASes have also introduced error into census tabulations, the 2020 DAS is the first to perturb {total} population counts in every geography smaller than states. This is necessary for the formal guarantees sought by the Census Bureau.
Relative to the population, these errors can be quite large in very small geographies (24.79\% in urban census blocks) but are much smaller even in moderately-sized geographies (0.31\% for counties of less than 1,000 people).

Many viewed the 2020 DAS as undermining the legitimacy of the data's use for redistricting and called for a rethinking of disclosure avoidance in the census~\citep{kenny2021use, Schneider_2021, Wezerek_Riper_2020, banks2025censusinvestigation}. Others disagreed \citep{cohen2022private, 2023jarmin}.
The argument against the new DAS was made most forcefully by Kenny \textit{et al.} \cite{kenny2021use}. They analyze the impact of the 2020 DAS using an ensemble-based redistricting analysis: a modern, computationally-intensive approach to exploring properties of the space of reasonable redistricting plans in a given geography.
They conclude that ``nonrandom local errors can aggregate into substantively large and unpredictable biases at district levels''
and claim ``the added noise makes it impossible to follow the principle of One Person, One Vote, as it is currently interpreted by courts and policy-makers'' \citep{kenny2021use}.

The concern is easy to understand: two districts whose populations are equal according to the official 2020 Redistricting Data will very likely have different populations according to the confidential 2020 CEF. If such discrepancies are large or biased, they could impact electoral power. 

Sharp legal thresholds can amplify discrepancies, turning a continuous source of error into a discrete category change. 
For example, 10\% population deviation is the cutoff between a state legislative plan being presumed discriminatory or not.
Prior work found that about 5\% of plans generated for the Louisiana state senate using the 2020 Redistricting Data would have exceeded that population deviation threshold according to the 2020 CEF~\cite{kenny2021use}.

This is what Alabama and Kenny \etal~\cite{kenny2021use} are worried about: that plans created using the Redistricting Data could turn out to be illegal\nt{---potentially ``impeding'' or even ``inhibiting'' the ability to ``draw fair lines,'' as Alabama put it.} (We question the underlying legal theory, as discussed in Section~\ref{sec:discussion}.)
Despite prior work quantifying the magnitude and characteristics of the perturbations, little is known about the interaction with threshold redistricting requirements.

\subsection*{Research Questions}   
The goal of this paper is to better understand how perturbations from disclosure avoidance combine with sharp legal thresholds to impact redistricting.  
{What sorts of discrepancies arise and why? Can they be mitigated?}

To that end, we consider a conscientious redistricter who wants to minimize the impact of disclosure avoidance on their map, even if not legally required.
Our redistricter only has access to the officially published 2020 Redistricting Data (after disclosure avoidance), but wants their maps to comply with threshold redistricting requirements as measured using the confidential CEF data (before disclosure avoidance).\footnotemark~

\footnotetext{%
    As explained in Section \ref{sec:data}, we do not have access to CEF and must use a different dataset SWAP as a stand in. See Section~\ref{sec:limitations} for a discussion of how this does and does not affect our results.}

We ask whether \nt{there exists methods by which} the redistricter is able to achieve two goals, measured against the confidential CEF.
First, to produce plans balanced within \emph{One Person, One Vote} (OPOV) population tolerances.
Second, to produce plans with as many \emph{majority-minority districts} (MMDs) as possible.

We study both questions at the state legislative level, where the legal 
standing of such plans depend on sharp thresholds (see Section~\ref{sec:legal-background}). 
For OPOV, population deviations less than 10\% are presumptively constitutional; larger deviations are presumptively unconstitutional. And showing that a political geography admits some number of majority-minority districts---districts with at least 50\% of the population in the relevant minority group---is necessary to successfully overturn racially gerrymandered plans.
We focus our examination of MMDs on majority-Black districts, reflecting a common scenario in voting rights litigation and the largest racial minority in the USA.

\subsection*{Overview of Methods}
This paper employs \emph{ensemble analysis}. We algorithmically sample large collections of possible districting plans 
{and observe how statistics of our ensembles differ between our two data sources} 
(see Section~\ref{sec:background} for details).

To study the effect of the DAS, one would ideally compare the data with and without disclosure avoidance, namely the 2020 CEF and 2020 Redistricting Data. However, the CEF is confidential to those outside of the Census Bureau. We instead use two Census datasets which we call SWAP and DEMO as stand-ins for the 2020 CEF and Redistricting Data, respectively. 
We are using these datasets exactly as intended: The Census Bureau created DEMO to help stakeholders study the impacts of the 2020 DAS by comparing it to SWAP.
See Section~\ref{sec:background} for more information and Section~\ref{sec:limitations} for further discussion on the limitations of using these datasets.

Our high-level approach is simple, and is adapted from Kenny \etal \citep{kenny2021use}.
First, we generate an ensemble of many redistricting plans.  These plans are drawn to satisfy some constraint---or to maximize some objective---on the {DEMO} dataset (standing in for 2020 Redistricting Data). 
Second, across plans in the ensemble, we see how often the constraint is violated---or how the value of the objective compares---on the {SWAP} dataset (standing in for 2020 CEF).

We call such disagreements in measurements between DEMO and SWAP \emph{discrepancies}. We  quantify discrepancies in two important quantities of redistricting plans: the population-deviation, and the number of majority-minority districts. 

\subsection*{{Contributions and Main Findings}}
We investigate the discrepancies that occur when maps created using data perturbed by the 2020 disclosure avoidance system are measured against sharp threshold requirements using the unseen, unperturbed dataset.
Overall, we find that thresholding can amplify the impact of disclosure avoidance noise, 
{in ways that are well-captured by simple models and possible to account for.}
\nt{In particular, there exists a method to satisfy population balance requirements on SWAP using only the DEMO data with confidence.}
\nt{We also provide simple probabilistic models that help explain our empirical observations.}

Below is a summary of our contributions and some of our main findings.
\nt{Our quantitative findings are specific to our experimental design (e.g., state legislative districts, using precincts as building blocks);  drawing quantitative conclusions for other redistricting settings (e.g., smaller sub-state districts, incorporating different constraints or objectives) would require adapting our techniques.}
We discuss these and other important limitations in Section~\ref{sec:limitations}.

\begin{description}
    \item[For One Person, One Vote (Section~\ref{sec:popbal})] We extend the ensemble analysis in prior work~\cite{kenny2021use} from one state legislative geography to 93. As in the prior work, we observe significant discrepancies. 
    For example, among plans sampled with at most 5\% population deviation on DEMO, the \emph{discrepancy rate}---the fraction of plans exceeding 5\% deviation on SWAP---was at least 40\% in half of the legislative geographies we examined.\footnote{5\% deviation as measured using the method of \cite{kenny2021use} and others, which approximates the 10\% deviation threshold as measured by courts (see Section~\ref{sec:notation}).}

    \nt{At the same time, there exists a method to draw districts that can substantially reduce or eliminate observed threshold violations within our district-generation procedure.}
    The countermeasure is simple:
    use a population deviation threshold in the redistricting sampling algorithm that is slightly tighter than the 5\% policy target (i.e., applying an offset).
    Sampling with a 4.5\% limit yields a discrepancy rate of 0\% for over half of geographies examined. 
    Sampling with a 4\% limit does so for 90\% geographies examined.

    \item[For majority-minority districts (Section~\ref{sec:gingles})]
    We extend the ensemble analysis in prior work~\cite{kenny2021use} from one state legislative geography to 52. We observe significant discrepancies, but in the opposite direction than prior work (see Section~\ref{sec:related_work}). For example, among Georgia state house plans sampled using DEMO, 9\% have a different number of majority-Black districts on SWAP---almost always fewer. 

    When sampling plans using a method that maximizes the number of majority-Black districts (as is common in Voting Rights Act litigation), this jumps to 66\%. 

    \item[Simple models for understanding discrepancies]   
    In both cases, we give simple probabilistic models that help explain many of the observed phenomena, including the effectiveness of the mitigation proposed for achieving population balance and the impact of maximizing the number of majority-Black districts on the discrepancies that result.
    \nt{Our hope is that these models prove useful for thinking about the impact of disclosure avoidance noise in settings beyond our specific experimental design.}
\end{description}

\section{Legal Context}
\label{sec:legal-background}

\subsection{One Person, One Vote}
\label{sec:legal:opov}
The One-Person, One-Vote principle holds that any two people living in the same state should have about the same voting power.
This means that the total population of different districts must be balanced according to the decennial census data (despite its imperfections, see Sec.~\ref{sec:background:noisy-data}).

For Congressional districts,  ``districts [must] be apportioned to achieve population equality as nearly as is practicable'' (\emph{Karcher v.\ Daggett}).
No amount of population deviation is considered de minimis. 
In practice, many states balance congressional districts to within a single person in total population.
This is one reason that Congressional redistricting is not the focus of our study: there is no policy-relevant numerical target that a redistricter can hope to achieve after accounting for noise.

State-level legislative districts must
    also be balanced in total population, although the requirement is much less strict.
For state legislative districts, population deviations below 10\% ``will ordinarily be considered de minimis,'' while
``disparities larger than 10\% creates a prima facie case of discrimination'' (\textit{Brown v.\ Thomson} 1983). 
In practice, state legislative population deviations are often much smaller than 10\%.

Note that courts define population deviation as the relative difference between the largest and smallest districts. We instead adopt the convention of \cite{kenny2021use} and others, using the maximum relative difference from the ideal district population. A 10\% deviation under the former definition is approximately equal to a 5\% deviation under the definition used in this paper (see Sec.~\ref{sec:notation}).

\subsection{Voting Rights Act}
Section 2 of the Voting Rights Act of 1965 outlaws vote dilution on the basis of race, including racial gerrymandering.
    The Supreme Court laid out a framework for judging racial gerrymandering claims in \emph{Thornburg v.\ Gingles} (1986), and recently reaffirmed the framework in \emph{Allen v.\ Milligan} (2023).

To successfully challenge an enacted map, a minority group must first pass a three-part threshold test: the Gingles preconditions.     
Our paper focuses on the first Gingles precondition, \emph{Gingles 1}. 
(The second and third preconditions together require voting to be polarized to such an extent that the minority group is prevented from electing its candidates of choice.)

Gingles 1 requires the minority group to be ``sufficiently large and geographically compact to constitute a majority in a reasonably configured district'' (\textit{Allen v.\ Milligan}, 2023). 
This is typically accomplished by producing for the court one or more demonstration maps (eleven in \emph{Allen}), each with more \emph{majority-minority districts} than the enacted map being challenged.
Gingles 1 is satisfied if at least one demonstration map is reasonably configured (``if it comports with traditional districting criteria'').

Throughout this paper, we focus on Black voters.
This reflects a common scenario in voting rights litigation and the largest racial minority in the USA according to Census data.
Hence, districts will be considered majority-minority if a majority of the \emph{voting age population} is Black (\textit{Bartlett v.\ Strickland}, 2009), excluding those who selected multiple races.

Gingles claims are also used to allege vote dilution against other minority groups. We present corresponding results for majority-Hispanic districts in Table~\ref{tab:hisp} in Appendix~\ref{app:methods}.

\subsection{Redistricting with Imperfect Data}
\label{sec:background:noisy-data}
It has long been recognized that the Decennial Census is far from perfect~\cite{zalesin2020beyond}. For example, coverage errors are large and vary by race: estimates for 2020 are 3.30\% Black undercount and 4.99\% Hispanic undercount, compared to 0.66\% White overcount~\cite{khubba2022national}. Even if the Decennial Census was {a perfect record of the population on Census Day}, populations can change significantly in the 10 years between censuses when districts have historically remained unchanged.

The Supreme Court recognizes these issues. Generally, the Census's official Redistricting Data is considered ``the only basis for good-faith attempts to achieve population equality'' despite being ``less than perfect'' (\textit{Karcher v.\ Daggett} 1983). 
But the Court has also acknowledged that deviating from the Redistricting Data can be appropriate. 
In \emph{Mahan v.\ Howell} (1973), a plan was considered malapportioned when about 18,000 people were known to live off of the naval base in which they were counted in the 1970 Census.

Alabama's lawsuit against the 2020 DAS raises an important question. What data should redistricting law treat as the ground truth: the data as  collected, edited, and imputed (the CEF) or as published (the Redistricting Data)?
No court has ever faced this precise question.

Kenny \textit{et al.} \citep{kenny2021use} consider Alabama's view---treat CEF as ground truth for redistricting---to be the law as ``currently interpreted by courts and policy-makers.''
\nt{Under this view, discrepancies between the Redistricting Data and the CEF could plausibly \emph{impede} redistricting by forcing mapmakers to account for measurement error, and might even be argued to \emph{inhibit} it if no plan could satisfy threshold requirements when judged against the CEF. (Looking ahead, our results suggest that for the state-legislative setting and plan-generation procedures we study, the ``impede'' concern is real but manageable; see Section~\ref{sec:discussion} for further discussion).}

We agree instead with Cohen \textit{et al.} \citep{cohen2022private}: the Redistricting Data should remain the only benchmark for redistricting requirements, considering  the overall scale of the noise from disclosure avoidance in the face of other sources of imperfection that the Supreme Court has long countenanced, and  the infeasibility of enforcing requirements measured against a confidential data source.
\nt{Under this view, redistricters would not be required to account for data discrepancies to meet legal threshold requirements, though they may still wish to do so as a prudential matter.}

\section{Methods and Data}
\label{sec:background}

\subsection{Data Sources} 
\label{sec:data}
We use data from the 2010 Census P.L. 94-171 Redistricting Data Summary Files (henceforth SWAP) and the Privacy-Protected 2010 Census Demonstration Data Vintage 2021-06-08 (DEMO) \citep{demodata}. 

SWAP is the official redistricting data from the 2010 Census, produced by applying the swapping-based 2010 DAS to the 2010 CEF. 
DEMO was produced by applying the 2020 DAS to the same 2010 CEF.     
Both DEMO and SWAP contain tabulations of population by voting age, sex, race, ethnicity for each geographic unit on the census geographic hierarchy---census blocks, block groups, tracts, counties, states, and the nation as a whole. 

SWAP and CEF agree on the total and voting age populations of every census geography, and agree on all statistics at the state level. 
In contrast, DEMO and CEF agree on the total population only at the state level, and all sub-state counts may differ. 
Neither SWAP nor DEMO agree with the CEF on population by race for census blocks.

To create the input data for our redistricting algorithms, we link these data with TIGER/Line Shapefiles \citep{tiger} and block equivalency files for voting precincts and legislative districts enacted after the 2010 Census \citep{equivalencyfiles}. For precincts, we additionally joined our data with dual graphs provided by the maintainers of the GerryChain software package~\cite{{gerrychain_julia_4649464}} for simpler integration into our algorithm.

\subsection{Ensemble Analysis}
\label{sec:ensembles}

In the past decade, algorithmic methods for studying gerrymandering and redistricting have made their way from academia \cite{thicket, CIRINCIONE2000189, chen2013unintentional, duchin2019locating, fifield2020essential} to the Supreme Court (\emph{Gill v.\ Whitford} (2018), \emph{Rucho v.\ Common Cause} (2019), \emph{Harper v.\ Moore} (2022), \emph{Allen v.\ Milligan} (2023)).\footnote{Ensemble-based evidence has been more persuasive at trial courts than at the Supreme Court. It was received positively by the dissent in \emph{Rucho v. Common Cause}, but less so in \emph{Allen v.\ Milligan}: ``[C]ourts should exercise caution before treating results produced by algorithms as all but dispositive of a \S2 claim.''}

Markov Chain Monte Carlo (MCMC) methods enable sampling ensembles of thousands or millions of redistricting plans from predefined distributions for analysis \citep{deford2022random,autry2020multiscale,fifield2020automated}.
Our paper makes use of the \recom algorithm, 
a merge-split MCMC algorithm which has been used by expert witnesses in 
gerrymandering cases to generate ensembles \citep[\textit{Harper v.\ Hall} 2022, \textit{Allen v.\ Milligan} 2023]{recom}. 
We use the implementation from the GerryChain software package \citep{gerrychain_julia_4649464} (with some modifications, as described in Section~\ref{sec:gingles}).

We use voting precincts as the smallest unit of geography for drawing districts.\footnote{We use the term precinct instead of the more correct ``voting district'' or VTD to avoid confusion with the much larger state and Congressional districts that are the main subject of this paper \citep{united1994voting}.}
Keeping precincts intact is often considered desirable in practice by state legislatures and previous academic work on redistricting ensembles has used precincts to compose districts \cite{kenny2021use}. For a few states, precinct data was incompatible with our experimental setup (either due to especially large precincts or incomplete geographical data). In those cases, we used block groups as the building block of districts in our analyses. See~Section~\ref{sec:limitations} for a discussion of how these choices may have affected our results.

\paragraph{Details for population balance (Section~\ref{sec:popbal})}
Except where otherwise noted, each ensemble contains 100,000 plans generated by running \recom for 1,000,000 steps and selecting every tenth plan.
We impose various population deviation limits on the MCMC sampler, as described in Section~\ref{sec:popbal}.

We generate ensembles for each of 93 state legislative geographies. 
These include the upper and lower geographies for 45 bicameral states and one geography for Nebraska's unicameral legislature. It excludes lower houses in 4 states (Hawaii, New Hampshire, North Dakota, and Vermont) and upper houses in 2 states (Hawaii and North Dakota) where our MCMC chains do not find any plans that are population balanced to within 5\% in DEMO. This limitation is due to our use of voting precincts rather than Census blocks as the unit of geography in our chains. Since the state legislative districts in these states are relatively small in population, it is computationally inefficient to find valid districts composed of precincts using MCMC methods.

\paragraph{Details for majority-minority districts (Section~\ref{sec:gingles})}
In Section \ref{sec:gingles}, we generate plans using both \recom and a modified version of \recom optimized to select 
plans with more majority-minority districts (MMDs). The latter uses a technique called 
\emph{short bursts}, which has been shown to outperform other techniques including biased random walks and simulated annealing \citep{cannon2023voting}. Details are deferred to Section~\ref{sec:shortburst}.

Ensembles using unmodified \recom contain 100,000 plans generated as above. 
Ensembles using short-bursts optimization contain 1,000 plans each (except for the Georgia state house ensemble which contains 5,000 plans).

Of the 93 state legislative geographies described above, our ensembles found at least one majority-Black district in 52 of them (27 lower house, 25 upper house). We restrict our analysis of MMD redistricting to these 52 geographies. 
For each geography, we impose a population deviation limit of 5\% minus the \emph{critical offset} found in Section~\ref{sec:popbal}.

\paragraph{Convergence}
Table~\ref{tab:convergence} in the Appendix summarizes results for Markov chain convergence tests for our main ensemble analyses.
Using the Gelman-Rubin split-$\hat{R}$ diagnostic \cite{gelman2014bayesian}, we find that estimates of the metrics we study in Section~\ref{sec:popbal} show clear signs of convergence in the large majority of the geographies we study ($\hat{R} \leq 1.01$, $\mathsf{ESS} \geq 400$). 
Compared to the analyses in Section~\ref{sec:popbal}, fewer ensembles in Section~\ref{sec:gingles} exhibit clear signs of convergence. In that context, we view the convergence tests as less important: we are using MCMC sampling more as a way to optimize quantity of interest (the number of MMDs) rather than to estimate one (the MMD discrepancy), in line with the \emph{Gingles 1} test.

\subsection{Notation}
\label{sec:notation}
Let $\data$ denote a reference dataset, either DEMO or SWAP.
A state legislative redistricting plan $\plan$ is a partition of a state into $\numDist$ districts: $\dist_1$, \dots, $\dist_\numDist$.
Each district $\dist_i$ is a contiguous collection of census blocks.\footnote{\nt{In our main experiments, districts are constructed from larger units called \emph{precincts} (each precinct aggregates many census blocks); we use ``blocks'' here only to align with the Census geographic hierarchy. See Section~\ref{sec:limitations} for discussion of how using precincts (rather than blocks) may affect the observed impact of DAS noise.}}
We denote a district's population in $\data$ as $\pop_\data(\dist)$,
 and its voting-age population $\vap_\data(\dist)$.

The ideal population of each district in a plan is $\idealPop_\data = \frac{1}{\numDist}\cdot\sum_{i=1}^\numDist \pop_\data(\dist_i)$. 
It  depends only on the total population of the geography and the number of districts. 
Because the 2010 and 2020 DASes do not affect a state's total population, $\idealPop_\ddp = \idealPop_\sfo$ for state legislative redistricting. As this setting is our focus, we will use $\idealPop$ throughout.
Fixing $\idealPop$, we define the \emph{population deviation} of a district and of a plan, respectively, as
$\dev_\data(\dist) = |\pop_\data(\dist) - \idealPop|/\idealPop$ and $\dev_\data(\plan) = \max_{i=1,\dots,k}\,\dev_\data(D_i)$.
This follows the convention of \cite{kenny2021use} of measuring deviation relative to the ideal population, and is a measure directly supported by our MCMC sampler.\footnotemark~
    \footnotetext{\label{note:deviation}Courts typically instead measure deviation as the largest population minus the smallest, divided by the smallest. To guarantee that this latter measure is at most $\tau^*$, it suffices that $\dev(\plan) \le 2\tol^*/(2+\tol^*)$ (e.g., for $\tol^* = 10\%$, $\dev(\plan) \le 0.094$ suffices).}

We denote by $\bpop_\data(\dist)$ and $\bvap_\data(\dist)$ the total and voting-age  Black populations, respectively.
We say a district is \emph{majority-Black} if $\bvap_\data(\dist)/\vap_\data(\dist) >0.5$. The number of majority-Black districts in a plan $\plan$ according to $\data$ is denoted $\mmd_\data(\plan)$.

\section{{Related Work}}
\label{sec:related_work}

Two prior studies have used ensemble methods to study the effects of the 2020 DAS on redistricting, specifically on population balance and the analysis of racial gerrymandering \citep{kenny2021use,cohen2022private}. We discuss them below.
Other research on the 2020 DAS include studies of the impacts on different policy issues, including minority representation \citep{christ2022differential}, misallocation of federal funds \citep{steed2022policy}, and public health monitoring \citep{krieger2021impact}, for example. 

The Census Bureau has also published its own analyses of the 2020 DAS. 
The most relevant examines the impact of the 2020 DAS on the reliability and consistency of measurements of demographic characteristics in various geographies \citep{wrightirimata}.
Among other findings, it shows that the 2020 DAS changes the apparent fraction of the population belonging to the largest demographic group by at most 5\% in 95\% of block groups with total population 450 to 499. 
Such reliability guarantees are important for Voting Rights Act enforcement, but don't do away with issues presented by sharp thresholds, as is the focus of our work.

\subsection{Kenny \etal~\cite{kenny2021use}}
 Kenny \etal \citep{kenny2021use} study the effects of the 2020 DAS on redistricting through a variety of case studies using ensemble-based comparisons of DEMO and SWAP.\footnote{Only a non-final version of DEMO (called  ``DAS-12.2 data'' in \cite{kenny2021use}) was available at the time of their initial experiments. Results using the more recent DEMO data (``DAS-19.61 data'') are in~\cite[Supplementary Materials]{kenny2021use}.}
We adopt their basic setup: generate many maps using the DEMO data, and measure their characteristics on the SWAP data.

\paragraph{Population balance}
Kenny \etal study of the impact on population balance for state legislative districts, with the Louisiana state senate as a case study~\cite{kenny2021use}. They generate ensembles of 5,000 plans using a merge-split MCMC algorithm (the same algorithm we use, but a different implementation~\cite{redist}).
They find that a significant fraction of redistricting plans satisfying a given maximum population deviation $\tau$ on DEMO exceed the $\tau$ limit on SWAP~\citep[Figure S4.4]{kenny2021use}. 
For example, for $\tau = 1\%$, about $75\%$ of plans exceed the deviation limit on SWAP; for $\tau = 5\%$, about 5\% of plans do.
We include these results in~Figure~\ref{fig:invalid-by-tol-offset-LA}.

In Section~\ref{sec:popbal}, we greatly extend this analysis. We extend it in scale: using 93 state legislative geographies (instead of 1) and ensembles of 100,000 plans (instead of 5,000). We also extend it in scope: introducing \emph{offsets} to mitgate discrepancies between DEMO and SWAP and measuring their effect and also proposing a model to explain the empirical observations.

\paragraph{Majority-minority districts}
Kenny \textit{et al.} \citep{kenny2021use} consider how the DAS affects the apparent number of majority-minority districts that can be drawn~\cite[Table S4.2]{kenny2021use}, through a case study reexamining \emph{NAACP of Spring Valley v. East Ramapo Central School
District} (2020). They sample 10,000 plans (5\% max deviation) for the school district using DEMO, and count the number of MMDs in each plan based on DEMO and SWAP. The find  {fewer} majority-minority districts in DEMO versus SWAP.

We find the opposite, across many state legislative geographies (Section~\ref{sec:gingles}).    
As we explain next, we believe the disagreement stems from differences in how our two works \nt{(i)} define and count MMDs, \nt{and (ii) generate ensembles of redistricting plans}.

We define an MMD as a district having more than 50\% of \emph{voting age population} in the minority group. This is the measure used in Supreme Court cases including \emph{Allen v.\ Milligan} (2023), \emph{Bartlett v.\ Strickland} (2009), and others.
Kenny \etal instead use \emph{registered voters} as was done in the {Ramapo} case. 
Race data for registered voters in Ramapo are not available. To get around this, Kenny \etal count MMDs using inferred race derived from the surnames and addresses of registered voters using a technique called Bayesian Improved Surname Geocoding (BISG).
BISG predicts the race or ethnicity of the registered voters in a district based on their surname and the demographic makeup of the district as measured using census data (i.e., DEMO or SWAP). 
Because of this, the effect of the 2020 DAS on counting MMDs as measured in \cite{kenny2021use} is mediated through its effect on BISG. The latter is a  significant confounder: comparing DEMO and SWAP, the inferred ``proportions of Black and Hispanic [registered voters] are much smaller, especially in the blocks where they form a majority group''~\cite{kenny2021use}. 

\nt{The second} difference is our use of an optimization technique designed to sample plans with many majority-Black MMDs. Compared to the race-blind sampler used in~\cite{kenny2021use} and in Section~\ref{sec:popbal}, our MMD-optimizing sampler appears to exacerbate the discrepancies we observe.

\subsection{Cohen \etal~\cite{cohen2022private}}
Cohen \textit{et al.} \citep{cohen2022private} also use ensemble methods to study the effects of the 2020 DAS on redistricting. 
(But as we elaborate below, the methods used in that paper make it less directly relevant to the present work than it may seem.)
Like~\cite{kenny2021use}, Cohen \etal offer a detailed study of a small handful of case studies. Our study of offsets in Section~\ref{sec:popbal} was in part inspired by the proposal in \cite{cohen2022private} to tweak a standard method for measuring racially-polarized voting (the 2nd and 3rd \emph{Gingles} preconditions).

While it goes a long way to understanding the scale and characteristics of the noise introduced by disclosure avoidance, 
the prior work does not study the interaction between the noise from disclosure avoidance and the sharp thresholds. Perhaps this is one reason that \cite{cohen2022private} and \cite{kenny2021use} arrive at opposite conclusions about whether the official Redistricting Data should be treated as ground truth for legal tests.

Finally, Cohen \etal overlook the possibility that redistricting with the perturbed data might induce additional effects. 
Whereas our MMD analysis creates plans using the perturbed data, the closest case study in Cohen \etal~\cite{cohen2022private} only examines the demographic makeup of four existing districts in Navajo County, Arizona.

\paragraph{Methodology}
Cohen \etal do not use the Census demonstration data whatsoever---unlike the majority of independent analyses of the 2020 DAS's effects, including~\cite{kenny2021use} and ours.
Instead, they analyze samples of noised data generated using an early version of the 2020 DAS, which they call \emph{TopDown18} and that they run themselves. 
In brief, they run TopDown18 with various parameter settings and, separately, sample redistricting plans using various map drawing methods. Then they analyze the scale and characteristics of the noise from TopDown18, aggregated up to the sampled districts, for the parameters and map drawing methods tested.

One limitation of this approach is that TopDown18's algorithm, implementation, and parameters are very different than the final version of the 2020 DAS.
The other important limitation is the need to resort to using less-than-perfect data as input to the DAS, the real data being confidential. Cohen \etal used a reconstructed version of the 2010 Census data constructed to be consistent with the official tabulations published in SWAP~\cite{cohen2022private}. There is no way to know how the reconstructed dataset differs from the real data, or how those differences might have affected the analyses in~\cite{cohen2022private}.

\section{Balancing District Populations with Noise}
\label{sec:popbal}
This section investigates how perturbations from disclosure avoidance interact with population balance requirements.
We consider a data user's ability to use the public DEMO dataset (after disclosure avoidance) to generate redistricting plans satisfying a maximum population deviation target under the confidential CEF dataset (before disclosure avoidance). Because SWAP and CEF agree on population totals, we can use SWAP as a perfect stand-in for CEF.

A significant fraction of redistricting plans satisfying a given population 
deviation limit on DEMO exceeds that limit on SWAP, as first shown by Kenny \textit{et al.} \citep{kenny2021use}. 
They conclude that it is impossible to produce redistricting maps that adhere to One Person, One Vote requirements on the unseen SWAP dataset, pointing to state legislative districts as particularly challenging.

This section challenges that conclusion. 
We propose a straightforward method for ensuring that state legislative maps meet population balance thresholds under both datasets.
Our method---requiring a smaller population deviation on the public DEMO dataset---is very effective, and does not greatly increase the difficulty of finding valid plans\footnote{As evidence of the fact that decreasing this threshold does not greatly increase the difficulty of finding valid plans, all of the ensembles analyzed in Section~\ref{sec:gingles} are sampled using population tolerance offsets derived in Section~\ref{sec:popbal}.} nor require the adoption of new redistricting techniques.
See Section~\ref{sec:limitations} for a discussion of limitations.

We focus on state legislative districts for a few reasons. First, some population deviation is allowed but precise thresholds are still observed. Our techniques would be ill-suited for studying Congressional maps, where population deviations greater than 1 person must be justified, because our algorithms are unable to generate ensembles of maps such strict requirements.
Second, states offer enough diversity to observe meaningful variation, while also being few enough to cover exhaustively. Third, we are building on the existing case study of the LA state senate in~\cite{kenny2021use}.

\subsection{Meeting Population Deviation Limits Using Offsets}
\label{sec:offsets}

We begin with a case study of the Louisiana state senate---where the sort of population discrepancies we study were first observed by Kenny \textit{et al.} \citep[Figure S4.4]{kenny2021use}---then extend the analysis to 93 state legislative geographies.

We propose a simple approach: draw the plan to satisfy a slightly tighter limit $\tol - \offset$ on DEMO, for some value $0\le \offset \le \tau$. We call $\offset$ the \emph{offset}.
To evaluate this approach, we measure the \emph{discrepancy at $\tol$ with offset $\offset$}: the fraction of plans that exceed $\tol$ deviation under SWAP among an ensemble of plans with deviation at most $\tol-\offset$ under DEMO.
When $\offset = 0$, we call this the \emph{discrepancy at $\tol$} or the \emph{no-offset discrepancy rate}. 
Finally, we also consider what we call the \emph{critical offset}: the smallest offset with 0 discrepancies observed in our ensembles of 100,000 plans
(see Appendix~\ref{app:methods} for more information on the computation of the critical offset).

\nt{Conceptually, this is an existence exercise: for our district-construction procedure, does there exist an offset $\offset$ such that plans drawn to satisfy $\tau-\offset$ on DEMO are empirically likely (in our ensembles) to satisfy $\tau$ on SWAP. The resulting critical offsets are therefore specific to our simulation method and choice of geographic building blocks (e.g., precincts versus blocks), and should not be treated as universal constants for all redistricting processes (Section~\ref{sec:limitations}). Still, this approach can provide a useful framework for reasoning about the effects of disclosure avoidance on OPOV concerns.}

\subsubsection{A Case Study: The Louisiana State Senate}

\begin{figure}[h]
    \centering
    \includegraphics[width=\linewidth]{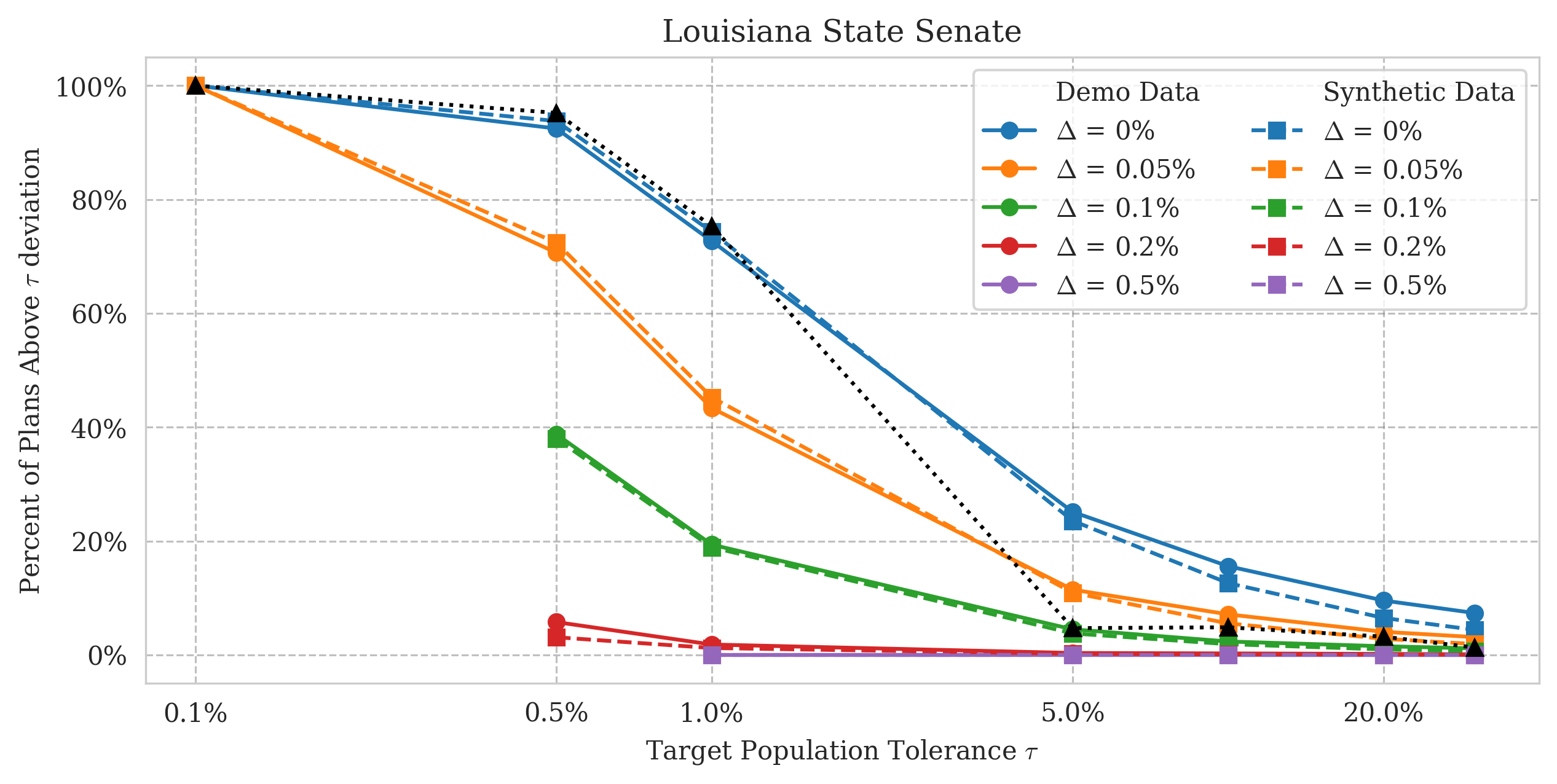}  
    \caption{\textbf{Discrepancy with offsets in the Louisiana state senate.} We plot the fraction of plans exceeding intended population tolerance limit $\tau$, with various offsets $\offset$ for the Louisiana state senate (i.e., discrepancy at $\tau$ with offset $\offset$).  Dots (solid lines) are computed using the DEMO and SWAP datasets with ensembles of 100,000 plans for each $\tau$ and $\offset$ (see Sec.~\ref{sec:offsets}). Squares (dashed lines) are computed from 100,000 samples from the statistical model of district populations and disclosure avoidance noise described in Sec.~\ref{sec:why-offsets-work}. Black triangles (dotted lines) represent the results obtained by \cite[Figure S4.4]{kenny2021use}.}
    \label{fig:invalid-by-tol-offset-LA}
\end{figure}

Replicating \cite{kenny2021use}, we measure
the no-offset discrepancy for each of six population deviation thresholds $\tol$ between 0.1\% and 30\%. 
Namely, we generate an ensemble of 100,000 plans for the LA state senate with population deviation at most $\tol$ according to DEMO, and count what fraction of those plans exceed $\tol$ deviation according to SWAP.

The results are summarized by the solid blue line in Figure~\ref{fig:invalid-by-tol-offset-LA}. 
The overall conclusion remains the same as in \cite{kenny2021use}: many plans drawn to a satisfy $\tol$ population deviation using DEMO exceed that threshold on SWAP. 
The exact numbers differ, perhaps due to differences in our experimental setups, such as a different MCMC sampler implementation and larger ensembles.

To test the effectiveness of offsets as a mitigation, we measure the discrepancy at $\tau$ and varying the offset parameter $\offset$, for the same six values of $\tol$ as before. 
Additionally, we compute the critical offset for the Louisiana State Senate for $\tau = 5\%$ to be $\offset = 0.33\%$ (averaged over 10 runs).

The solid lines in Figure~\ref{fig:invalid-by-tol-offset-LA} illustrate the results for $\offset = 0.05\%$, 0.1\%, 0.2\%, and 0.5\%. 
Using $\offset = 0.1\%$ reduces the discrepancy at $\tol =5\%$ from 23.4\% to 4.0\%. Using $\offset = 0.2\%$ reduces  the discrepancy at $\tol=1\%$ from 74.2\% to 1.1\%.  
Notably, the discrepancy at $\tol =5\%$ with offset $\offset = 0.5\%$ was 0\%. 
That is, of 100,000 plans generated using a 4.5\% tolerance on DEMO, none exceeded a 5\% tolerance on SWAP. 

\subsubsection{Offsets for State Legislative Redistricting}
We perform a similar analysis for 93 state legislative geographies.
For each geography, we measure the discrepancy at $\tol=5\%$ as we increase $\offset$ in steps of 0.05\% (generating an ensemble of {100,000} plans for each offset). For each geography we record the no-offset discrepancy ($\offset = 0\%$) and the critical offset (where the discrepancy first falls to 0\%). 

Table~\ref{tab:prop_invalid_all_states} gives a coarse summary of the results (Table~\ref{tab:offset} in Appendix~\ref{app:methods} gives much more detail).
Observe that the no-offset discrepancy rates vary widely. In the worst case (Kansas State House of Representatives) 96\% of plans exceeded 5\% population deviation under SWAP. But offsets provide a powerful mitigation: for most geographies, 
$\offset \le 0.75\%$ suffices to eliminate plans that exceed a 5\% population tolerance from our ensembles. 

\begin{table}[h]
    \centering
    \begin{tabular}{lrrrrr}
\toprule
                  &   Min &   $25^{th}$ &   $50^{th}$ &   $75^{th}$ &     Max \\
\midrule
 Discrepancy Rate & 4.73\% &      22.65\% &      40.59\% &      66.13\% & 100.00\% \\
 Critical Offset  & 0.10\% &       0.28\% &       0.44\% &       0.74\% &   2.18\% \\
\bottomrule
\end{tabular}
    \caption{Summary of discrepancy rates ($\offset = 0$) and critical offsets for 93 state legislative ensembles
and $\tau = 5\%$ (extrema and quartiles). Note the ensemble maximizing discrepancy does not maximize the
critical offset; likewise for the other order statistics.}
    \label{tab:prop_invalid_all_states}
\end{table}

\begin{figure}[h]
    \centering
    \includegraphics[width=\linewidth]{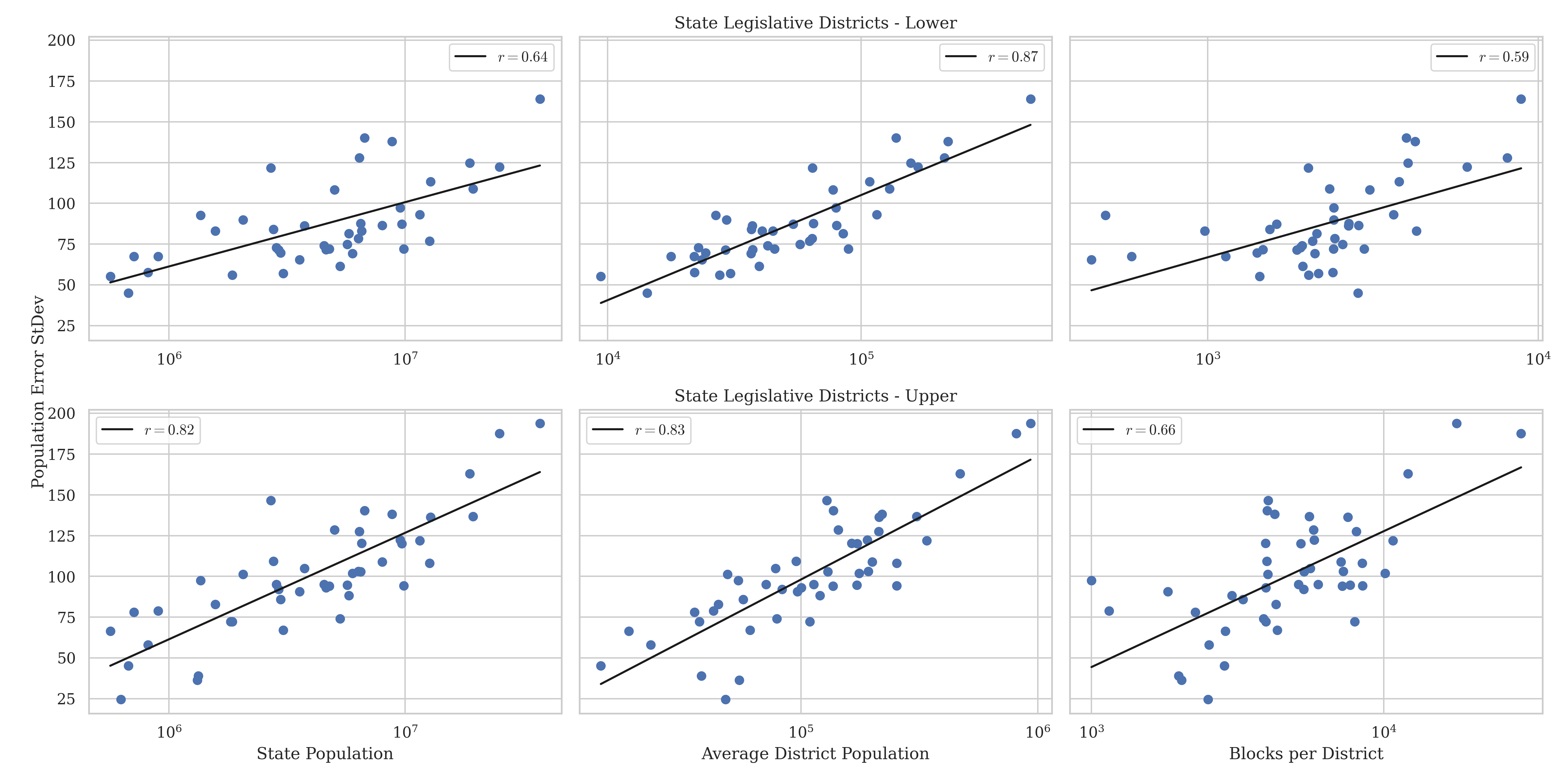}
    \caption{\textbf{Standard deviation of population error across state legislative geographies.} Standard deviations of the population error ($\pop_\ddp(D) - \pop_\sfo(D)$) over districts in ensembles generated for state upper and lower legislative districting plans. Each point represents a state legislative geography. 
    The left column plots the logarithm of each state's population on the horizontal axis. The center column plots the logarithm of each legislative geography's average district population on the horizontal axis. The right column plots the logarithm of the total number of blocks in each state divided by the number of districts in each respective legislative geography on the horizontal axis.}
    \label{fig:variance_analysis}
\end{figure}

We do not know why different geographies exhibit different discrepancies and critical offsets, although the size of districts appears to play an important role. In Figure~\ref{fig:variance_analysis}, we examine the standard deviations of the observed district population errors in our state lower and upper legislative district ensembles. Notably, we observe a strong positive correlation between the logarithm of average district population and error standard deviation in both state lower houses ($r = 0.87$) and state upper houses ($r = 0.83$). We also observe weaker correlations between the logarithm of total state population and error standard deviation (lower houses: $r = 0.64$, upper houses: $r = 0.82$) and between the logarithm of the average number of census blocks per district and error standard deviation (lower houses: $r=0.59$, upper houses: $r=0.66$).

\begin{figure}[h]
    \centering
    \includegraphics[width=\linewidth]{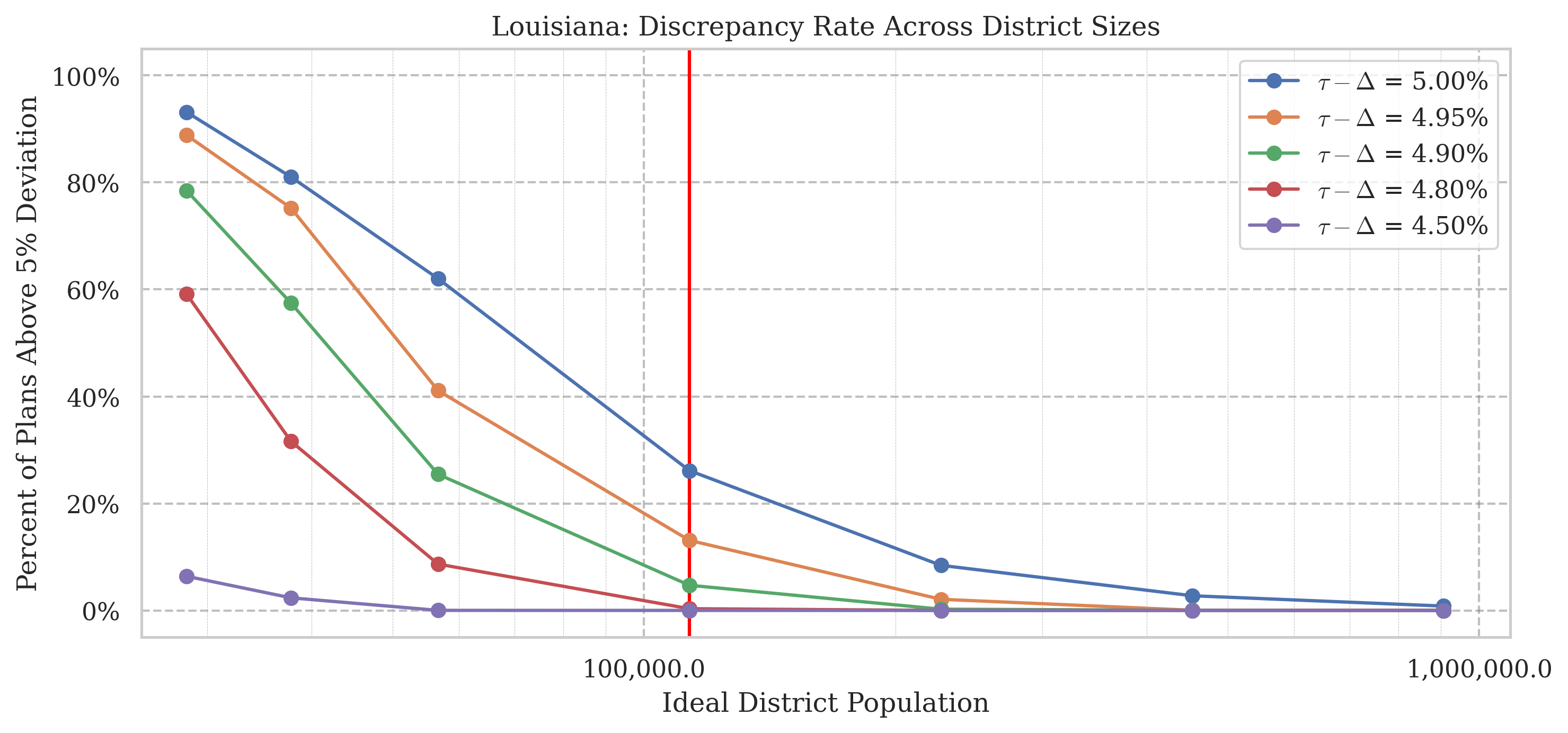}
    \caption{\textbf{Discrepancies by district size in Louisiana.} Measuring the discrepancy rate of Louisiana redistricting plans over various district sizes and population tolerance offsets. Each point represents an ensemble that was generated using a population tolerance threshold of $5\% - \offset$ on DEMO (for several values of $\offset$) and evaluated with a $\tau = 5\%$ threshold on SWAP. The ideal district populations correspond to plans with 5, 10, 20, 40, 80, 120, and 160 plans. The vertical red line corresponds with the actual district size of the Louisiana State Senate.}
    \label{fig:districtsize}
\end{figure}

We ran additional experiments on the state of Louisiana to understand the link between district size and discrepancy rate. For each value of $\offset$ we examined in Figure~\ref{fig:invalid-by-tol-offset-LA}, we measure the discrepancy rate at $\tau = 5\%$ for plans with different numbers of districts, ranging from 5 to 160. This corresponds to average district populations between 28,334 and 906,674 people.\footnote{For comparison: Louisiana Congressional districts contain about 766,000 people; New Orleans City Council districts contain about 72,500 people.} Results are shown in Figure~\ref{fig:districtsize}. We observe that for all values of $\tau - \offset$, the discrepancy rate is monotonically decreasing as the ideal district size increases.

\subsection{Why Offsets Work: Most Deviation isn't from Disclosure Avoidance}
\label{sec:why-offsets-work}

The effectiveness of our mitigation can be understood by disentangling the population deviation caused by disclosure avoidance from the deviation caused by other factors.
The population deviation of an individual district $\dev_\sfo(\dist)$ consists of two components:\footnote{This identity uses the fact that $\idealPop_\ddp = \idealPop_\sfo$ for state legislatures. Analyzing sub-state redistricting requires more care.}
\small
\[\dev_\sfo(\dist) = \biggl|
\underbrace{\frac{\pop_\ddp(\dist) - \idealPop}{\idealPop} }_{\text{signed }\dev_\ddp(D)}
+ 
\underbrace{\frac{\pop_\sfo(\dist) - \pop_\ddp(\dist)}{\idealPop}}_{\error(\dist)}\biggr|.\]
\normalsize
The first component is a signed version of $\dev_\ddp(\dist)$---the apparent deviation in DEMO. This is entirely under the control of the mapmaker. 
The second component is the additional error from disclosure avoidance, which we denote by $\error(\dist)$.

\begin{figure}[ht!]
    \centering
    \includegraphics[width=\linewidth]{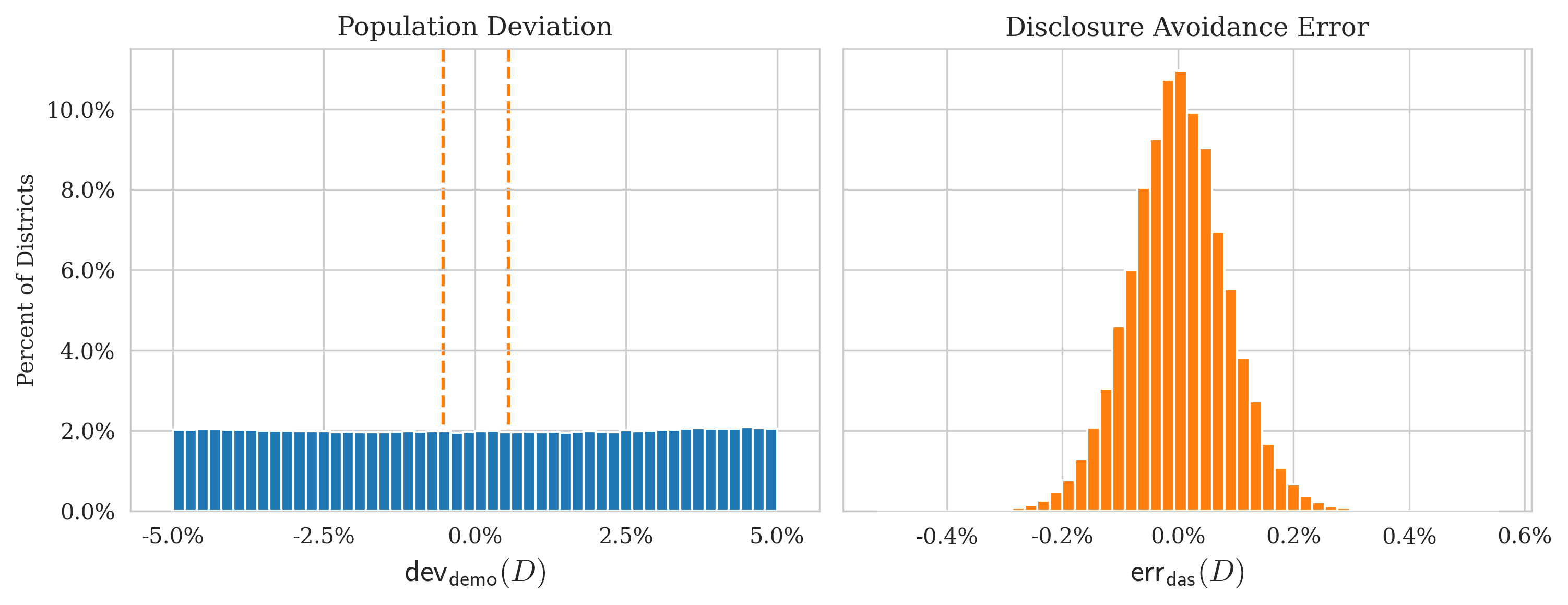}
    \caption{\textbf{The two components of population deviation in districts.}  Histograms of apparent population deviation in DEMO ($\dev_\ddp(\dist) 
 = (\pop_\ddp(\dist) - \idealPop/\idealPop$) and the additional error from disclosure avoidance ($\error(\dist) = (\pop_\sfo(\dist) - \pop_\ddp(\dist))/\idealPop$)
 for each unique district included in an ensemble of Louisiana state senate plans sampled with a 5\% population tolerance on the DEMO data. Note the very different scales on the horizontal axis. Together, these terms make up the population of a state legislative district's population deviation. The dashed vertical lines behind the left histogram represent the maximum and minimum error values observed in the right histogram.}
    \label{fig:error_decomposition_LA}
\end{figure}

Plotting these two terms side-by-side clarifies their relative import. Figure~\ref{fig:error_decomposition_LA} plots histograms of signed $\dev_\ddp(\dist)$ and $\error(\dist)$ for all the LA state senate districts in our $\tol = 5\%$ ensembles. Note the different scales on the horizontal axes.
The signed $\dev_\ddp(\dist)$ is roughly uniform across the acceptable range $[-5\%,5\%]$. The DAS noise $\error(\dist)$ is highly concentrated, with mean 0.0\% and standard deviation 0.080\%. Only a small fraction of districts have $\dev_\ddp(\dist)$ close enough to $5\%$ for the DAS noise to matter. Using an offset moves  $\dev_\ddp(\dist)$ away from the 5\% threshold, further lowering the number of districts where the DAS noise matters.

Figure~\ref{fig:error_decomposition_LA} suggests a very simple probabilistic model for population deviations under SWAP.
The deviation of a plan is sampled as the maximum over district-level deviations: $\dev_\sfo(\plan) = \max_{i} \dev_\sfo(\dist_i)$.
Each $\dev_\sfo(\dist_i)$ is independently sampled as $|X + E|$, where $X$ is uniform over $[-(\tol-\offset), \tol-\offset]$ and $E\sim\normal(\mu,\sigma^2)$.
The parameters $\mu=0.0\%$ and $\sigma=0.080\%$ are the empirical mean and standard deviation of $\error(\dist)$ across districts in our LA state senate ensembles. 

The dashed lines in Figure~\ref{fig:invalid-by-tol-offset-LA}  show the fraction of 100,000 values of $\dev_\sfo(\plan)$ sampled as above that exceeded $\tol$, for each $\tol$ and $\offset$. 
Qualitatively, the model closely approximates the empirical data for the LA senate, suggesting this is a useful mental model for understanding the effect of offsets. 

Note that this is meant only as a simple, usable model of the phenomenon. In particular, the observed values of $\error(\dist)$ do not appear normally distributed, failing the parametric bootstrap goodness of fit test implemented in \texttt{SciPy}~\cite{scipy}.)

\subsection{A Closer Look at Errors from Disclosure Avoidance}

To understand the effect of the 2020 DAS on OPOV questions, it appears that we must understand $|\error(\dist)|$: the magnitude of population error in a district $\dist$ due to disclosure avoidance, as a fraction of the ideal district population. This is perhaps unsurprising. 

We briefly discuss this error in enacted districts, as compared to other sources of error, and in relation to racial biases in the 2020 DAS. 

\paragraph{Errors in enacted districts} 
Real redistricting plans are drawn by people, not sampled from ensembles. We do not know how offsets  fare in a real redistricting setting. 
But the distribution of $|\error(\dist)|$ on districts created by the political process can give us some initial idea. 
Using DEMO and SWAP, we compute $|\error(\dist)|$ for every congressional and state legislative district enacted after the 2010 Decennial Census (rather than for algorithmic ensembles of districts). We group the districts by ideal population, which doesn't depend on the redistricting plan.
Table~\ref{tab:real_dist_errors} reports the maximum, 98th, and 90th percentile values in each group.
For the overwhelming majority of geographies---including all with ideal population above $32,000$---the relative error magnitude from disclosure avoidance was well under 1\%. A Rhode Island State House of Representatives district saw the  largest observed relative error: 2.14\%, corresponding to just 301 people.

A redistricter can use these numbers to guide the use of offsets where population deviation under the CEF is a concern, though more research would be warranted. If the ideal district population is $\idealPop = 200,000$, an offset of $\offset = 0.51\%$ might be a conservative choice; if $\idealPop = 20,000$, an offset of {$\offset = 1.37\%$} appears more appropriate.

\begin{table*}
\centering

\begin{tabular}{lrrrrrrrrr}
\toprule
 Ideal population $\idealPop$   &     \ensuremath{<}8k &     8-16k &    16-32k &    32-64k &   64-128k &   128-256k &   256-512k &   \ensuremath{\geq}512k \\
\midrule
  Count & 104 & 641 & 1,057 & 2,010 & 1,392 & 1,232&   216 &    506 \\
 Max    &   0.0168 &   0.0214 &    0.0137 &    0.0069 &    0.0048 &     0.0051 &      0.0041 &   0.0011 \\
 98$^{th}$ pct               &   0.0088 &   0.0111 &    0.0063 &    0.0039 &    0.0026 &     0.0017 &      0.0015 &   0.0006 \\
 90$^{th}$ pct               &   0.0047 &   0.0064 &    0.0039 &    0.0023 &    0.0015 &     0.0010 &      0.0006 &   0.0003 \\
\bottomrule
\end{tabular}
\caption{Error from disclosure avoidance in state legislative districts and Congressional districts enacted after the release of the 2010 census redistricting data, as a fraction of ideal population: $|\error(\dist)| = |\pop_\ddp(\dist) - \pop_\sfo(\dist)|/\idealPop$. Districts are grouped by ideal population.}
\label{tab:real_dist_errors}
\end{table*}

\paragraph{Magnitude relative to population drift}
As a point of comparison for the population deviations so far,  we analyze district population imbalances arising from population shifts in the 10 years between decennial censuses. This is only one acknowledged source of population deviation among districts, and ignores systematic undercounts and overcounts that differ by race and ethnicity \citep{pes}.
Using data from IPUMS that standardizes congressional district geographies between decennial censuses, we compare total populations of congressional districts of the 110th-112th 
Congresses in both the 2000 and 2010 censuses \citep{nhgis}.\footnote{Because Census 
geographies (i.e. block boundaries) are modified every 10 years, 
districts drawn using 2000 census data may split 2010 census blocks, complicating 
comparisons between years.}
When measured with the 2000 data used to draw the districts, the deviations are relatively small. Of states with more than one Congressional district, the average deviation is just 613 people ($\sim$0.1\% of the ideal population), while the maximum is 6,698 ($\sim$1\% of the ideal). By the 2010 Census, the average increased to 137,974 people ($\sim$19\% of the 
ideal), while the maximum increased to 340,948 people (Texas, $\sim$43\% of the ideal).

\paragraph{Biases in population discrepancies}
Prior work found that noise from the 2020 DAS is correlated with racial/ethnic homogeneity \citep{kenny2021use, pettiflaxman}.
Specifically, precincts with higher Herfindahl–Hirschman index---a measure of racial and ethnic homogeneity---tend to have their populations inflated, and vice versa. \citep{kenny2021use}.
This motivates the following question: Does the distribution of district-level errors in our ensembles vary by the racial and ethnic makeup of the districts?
If so, it might systematically shift voting power among groups even if district populations are still within the acceptable bounds.
Fortunately, we do not observe any meaningful effect. For the 876,800 unique districts we sampled for the Georgia state house, a geography for which small changes in voting power of Black residents could have a meaningful impact on the makeup of the legislature, we find that the Pearson correlation between per-district population error and Herfindahl-Hirschman index is minute: $r = 0.07$, 95\% CI $(0.068, 0.072)$.

\section{Counting Majority-Minority Districts with Noise}
\label{sec:gingles}
To challenge an enacted electoral map for racial vote dilution, plaintiffs must meet the Gingles 1 precondition. Gingles 1 requires showing that one can draw more majority-minority districts (MMDs) than exist in the enacted map being challenged. 
This is usually done by submitting expert reports with one or more such illustrative plans. 

This section investigates how perturbations from disclosure avoidance interact with Gingles 1.
We consider a data user's ability to use the public DEMO dataset (after disclosure avoidance) to produce illustrative plans with many majority-minority districts under the confidential CEF dataset (before disclosure avoidance). 
Notice that these illustrative plans may be quite different than those produced in the normal course of redistricting: they must have more majority-minority districts to satisfy Gingles 1.
For this reason, we also investigate the additional impact of optimizing the number of majority-minority districts using a technique called \emph{short-burt optimization}~\citep{cannon2023voting} described below.

Unfortunately, the data needed to study these questions directly is unavailable: SWAP and CEF do not agree on population by race, though they do agree on voting age population.
Therefore, we will use SWAP as an imperfect stand-in for CEF. As a result, the discrepancies we observe reflect the combined impact of perturbations from the 2010 DAS (CEF$\to$SWAP) and the 2020 DAS (CEF$\to$DEMO).
We discuss this limitation further in Section~\ref{sec:limitations}, and take a first step towards overcoming it in Appendix~\ref{app:swap-swap}.

This section specifically focuses on majority-Black districts and uses MMD to mean a district where a majority of the voting age population is Black, excluding those who select multiple races. As before, we restrict our attention to state legislative redistricting.

\subsection{Short-Burst Optimization}\label{sec:shortburst}
Gingles 1 incentivizes plaintiffs to maximize the number of MMDs in the illustrative plan, shifting the distribution of plans.
To study the effect of this shift, we generate ensembles of plans using a technique called \emph{short-burst optimization}, which is designed to output plans with many majority-Black MMDs~\citep{cannon2023voting}.
It is generally intractable to determine the maximum possible number of MMDs in a plan at the scale of a US state.
Therefore, we use the optimized chains as a proxy for a redistricter who is trying to draw a plan with a near-maximal number of majority-minority districts. 

We use short-burst optimization in a manner suggested to us by Duchin (personal communication, August 2025).
We run 100 chains (500 for Georgia) with short burst optimization for 100,000 steps each in parallel. From each chain, we randomly subsample 10 plans from among those with highest MMD count in DEMO, resulting in 1,000-plan ensembles (5,000 for Georgia).

This approach is somewhat heuristic, but it allows us to observe the effects of incentivizing maps with certain characteristics, as Gingles 1 does.
The ensembles can also be seen as containing plans that might have been selected as a Gingles 1 illustrative plan by a redistricter who used short-burst optimization to find maps with many MMDs. While the human and the algorithm differ in their methods for selecting plans, this methodology sheds light on the qualities of plans that are drawn from a nearly-maximal distribution.

\nt{This approach is computationally intensive, requiring hundreds of CPU-hours to generate an ensembles of 5,000 plans for the state of Georgia. 
We ran fewer chains for other states in order to allow us to complete our experiments in a reasonable timeframe. 
We observed that analyzing the first 1,000 plans of the Georgia ensemble yielded similar results to the full 5,000 plan ensemble.}

\subsection{Measuring MMD Discrepancy}
\label{sec:mmd_A}

We begin with a case study of Georgia state house plans, then extend the analysis to 50 state legislative geographies.
The Georgia house serves as a good illustrative geography due to its significant Black population and large legislature. 

We examine the \emph{MMD discrepancy} of plans generated using the DEMO dataset: $\mmd_\ddp(\plan) - \mmd_\sfo(\plan)$. We measure the \emph{mean MMD discrepancy} and the \emph{non-zero discrepancy rate}---the fraction of plans with non-zero MMD discrepancy. 

\subsubsection{Georgia State House Plans}
\label{sec:GA-mmd}
\begin{figure}[h]
    \centering
    \includegraphics[width=\linewidth]{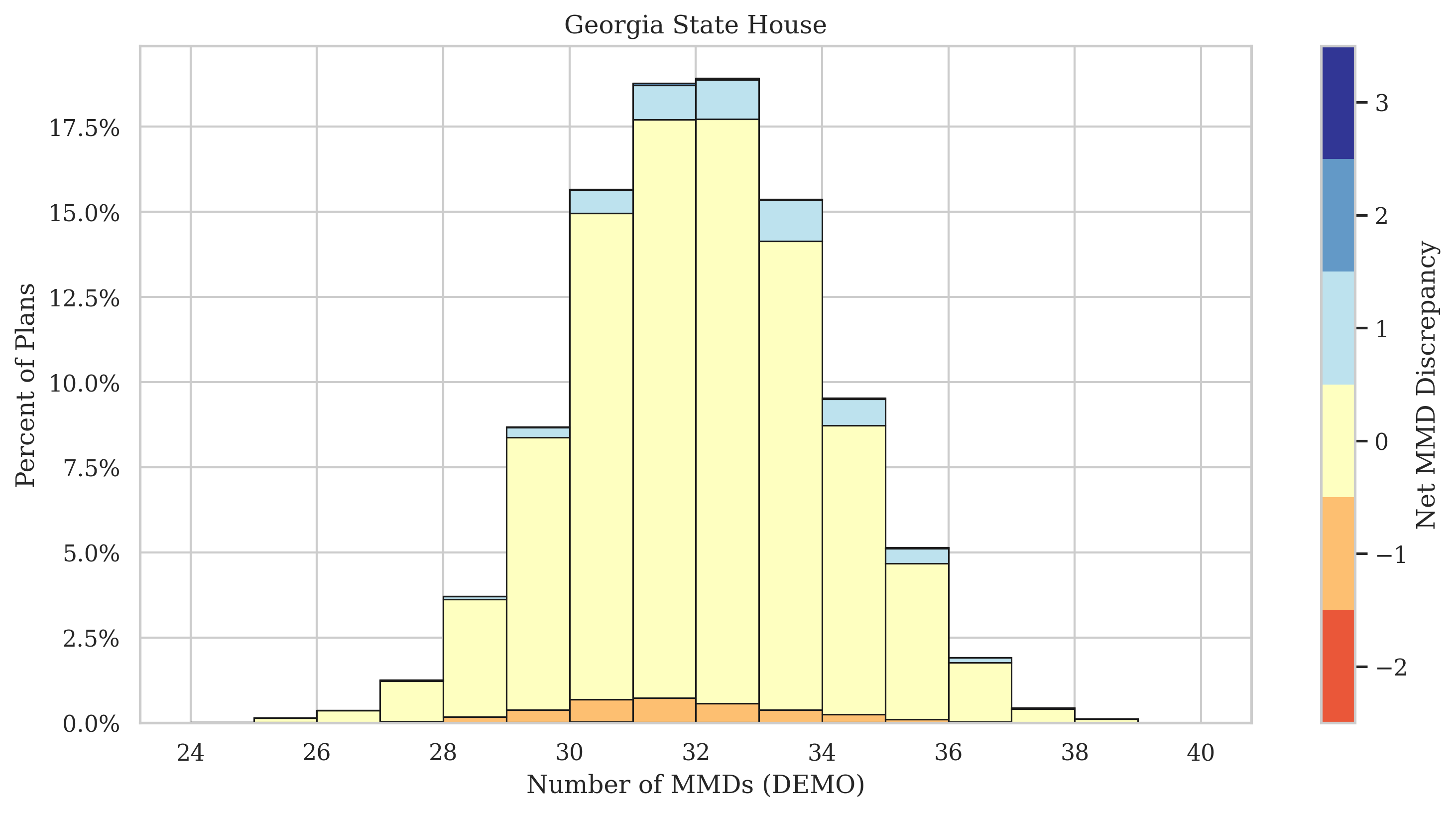}
    \caption{\textbf{MMD discrepancies in the Georgia state house \vanilla ensemble.} 
    We plot the number of majority-Black districts in plans sampled in our \vanilla ensemble.
    Plans are grouped according to the number of majority-Black districts as measured using DEMO. Bars are colored to indicate the fraction of those plans which have the corresponding MMD discrepancy. Each ensemble contains 100,000 plans sampled with a 4.4\% population deviation limit (utilizing $\offset = 0.6\%$) using the DEMO data. }
    \label{fig:ga_stacked_hist_vanilla}
\end{figure}

Figure~\ref{fig:ga_stacked_hist_vanilla} illustrates the MMD discrepancies that arise in 100,000 plans sampled using our base merge-split MCMC algorithm (Section~\ref{sec:ensembles}).
The histogram breaks down the plans by number of majority-Black MMDs according to the DEMO dataset. Each histogram bin is further broken down by MMD discrepancy: the difference between the number of MMDs as measured by DEMO and SWAP. 
Across the whole ensemble, the mean MMD discrepancy is 0.03, and 9\% of plans have a non-zero MMD discrepancy.
Furthermore, plans with more MMDs have slightly greater discrepancies (Pearson's $r=0.08$). 

\begin{figure*}[ht!]
    \centering
    \includegraphics[width=\linewidth]{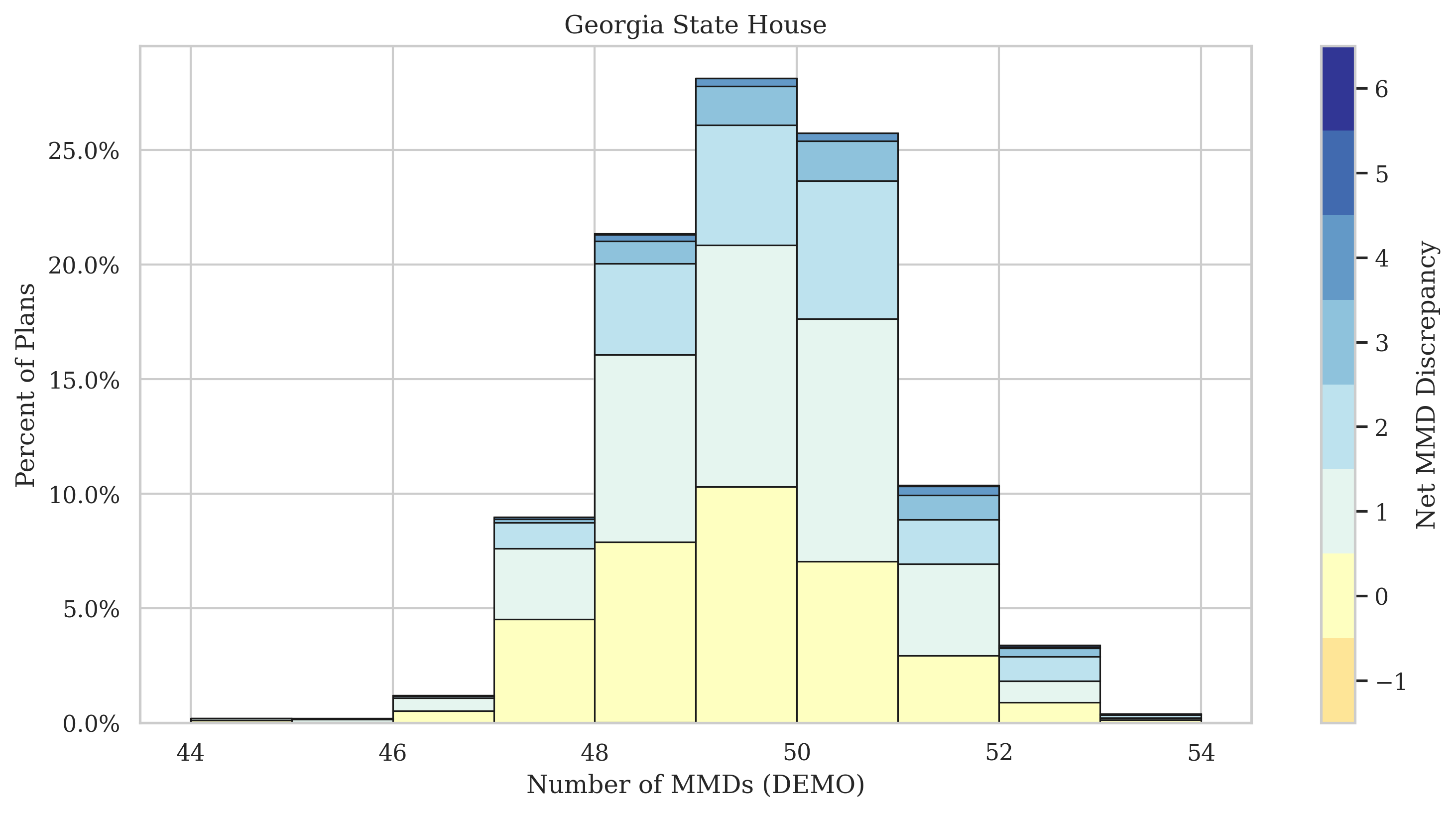}
    \caption{\textbf{MMD discrepancies in the Georgia state house optimized ensemble.} 
    We plot the number of majority-Black districts in plans sampled in our \bursts ensemble, which is designed to produce plans with many MMDs.
    Plans are grouped according to the number of majority-Black districts as measured using DEMO. Bars are colored to indicate the fraction of those plans which have the corresponding MMD discrepancy. Each ensemble contains 100,000 plans sampled with a 4.4\% population deviation limit (utilizing $\offset = 0.6\%$ offset listed in Table~\ref{tab:offset}) using the DEMO data. }
    \label{fig:ga_stacked_hist}
\end{figure*}

Figure~\ref{fig:ga_stacked_hist} illustrates the results of the same analysis for 5,000 plans generated using short-bursts optimization
These numbers are markedly higher: the mean MMD discrepancy is {1.02, and 66\%} of plans have a non-zero MMD discrepancy.
The correlation between number of MMDs and MMD discrepancy is also higher
(Pearson's {$r = 0.16$}). 

\subsubsection{Other State Legislative Geographies}
We examine the MMD discrepancies in the 52 geographies where our short burst ensembles found at least one MMD (27 lower house, 25 upper house). 
Full results are included in Table~\ref{tab:mmd_discreps} in Appendix~\ref{app:methods}.
Different geographies experience very different levels of MMD discrepancy. Some geographies experience very minor discrepancies; others very significant (e.g., 0.9\% of CA lower house plans have non-zero discrepancies, compared to 78.9\% of MS lower house plans). 

Still, some patterns we see in the GA house generalize to almost all geographies.
First, positive MMD discrepancies---more MMDs according to DEMO than SWAP---are both more frequent and larger in magnitude than negative discrepancies. Second, plans with more MMDs have greater discrepancies. Finally, we observe that MMD discrepancies are generally greater in lower houses than in the (typically smaller) upper houses within the same state.

\subsubsection{DEMO vs SWAP vs CEF}
\nt{
Recall that the discrepancies we observe reflect the combined impact of perturbations from the 2010 DAS (CEF$\to$SWAP) and the 2020 DAS (CEF$\to$DEMO). This is inherent to any analysis that compares the DEMO data to the 2010 Redistricting Data, including most analyses of the effects of the 2020 DAS.
(Additional results in Appendix~\ref{app:swap-swap} suggest discrepancies might be similar if we were to use the 2010 CEF as a baseline, though these results require strong assumptions on the re-implementation of the swapping algorithm in~\cite{Ballesteros_2025}.)

Still, characterizing the expected discrepancies between DEMO and SWAP can be useful for redistricters in practice. If one accepts the need for disclosure avoidance in census tabulations but is concerned that the new DAS yields data that is less fit for use than the previous DAS, our results should provide some reassurance.
Our results can be seen as useful for understanding whether DEMO is sufficiently close to SWAP for use in redistricting, setting aside whether SWAP is sufficiently close to the 2010 CEF to be fit for use. 
If it is, and if SWAP was fit for use for redistricting (which was never under question), then that lends support for DEMO's fitness for use as well.
}

\subsection{The Connection between BVAP Margin and MMD Discrepancy}
\label{sec:bvap-margin}

Above, when drawing districts with DEMO and measuring against SWAP we see: (1) a bias towards more MMDs in DEMO than SWAP, and (2) the bias increases as the number of MMDs increases within a geography.
This section \nt{describes a simple probabilistic model that helps explain these observations.}
\nt{We do so by connecting plan-level discrepancies to a district-level quantity (BVAP margin) whose sign determines whether a district is counted as majority-Black.}

\paragraph{BVAP margin}
The key quantity is \emph{BVAP margin}: the size of a district's Black voting age majority (i.e., Black voting age population minus half of the total voting age population).
We denote a district's BVAP margins under DEMO and SWAP as:
\begin{align}
    \bvapmargin_\ddp(\dist) &= \bvap_\ddp(\dist) - 0.5\cdot\vap_\ddp(\dist)\\
    \bvapmargin_\sfo(\dist) &= \bvap_\sfo(\dist) - 0.5\cdot\vap_\ddp(\dist) \notag
\end{align}
\nt{By definition}, a district is an MMD according to DEMO if and only if its BVAP margin under DEMO is positive: \nt{$\bvapmargin_\ddp(\dist)>0$}. Likewise, for SWAP.

\nt{Given only the DEMO dataset, a redistricter can reason about MMDs under SWAP by observing the following:}
\begin{equation}
    \bvapmargin_\sfo(\dist) = \bvapmargin_\ddp(\dist) + \underbrace{\bigl(\bvapmargin_\sfo(\dist) - \bvapmargin_\ddp(\dist)\bigr)}_{\error^\mmd(\dist)} \label{eqn:bvap-decomposition}
\end{equation}
\nt{This decomposition separates what is observable by the redistricter from what is not (as Section~\ref{sec:why-offsets-work} did for population balance).}
The first component $\bvapmargin_\ddp(\dist)$ is entirely under the control of the mapmaker. The second component is the change in BVAP margin from disclosure avoidance, which we denote $\error^\mmd(\dist)$.

\nt{By equation~\eqref{eqn:bvap-decomposition}, a district contributes to plan-level MMD discrepancy precisely when the disclosure-avoidance term $\error^\mmd(\dist)$ causes a \emph{sign flip} between $\bvapmargin_\ddp(\dist)$ and $\bvapmargin_\sfo(\dist)$. Districts with $\bvapmargin_\ddp(\dist)\approx 0$ are the most likely to flip.}
\nt{Therefore, we might hope to explain characteristics of the plan-level MMD discrepancies by looking at the characteristics of $\bvapmargin_\ddp(\dist)$ and $\error^\mmd(\dist)$, and focusing specifically on \emph{near-threshold} districts where $\bvapmargin_\ddp(\dist) \approx 0$.}

\paragraph{Understanding biases in MMD discrepancy}
In our experimental setup, we tend to see a bias towards more MMDs in DEMO than in SWAP. This corresponds to plans having more districts $D$ satisfying $\bvapmargin_\sfo(\dist)<0<\bvapmargin_\ddp(\dist)$ than the reverse. The more MMDs a plan has, the greater the observed bias tends to be.
\nt{Why might this be?}

\nt{The reasoning above suggests that we find an explanation by looking at biases in $\bvapmargin_\ddp(\dist)$ and $\error^\mmd(\dist)$, focusing on near-threshold districts where the BVAP margin is close to 0 (i.e., where sign flips / discrepancies are plausible).}

Hence, we consider two possible sources for the observed bias. 
\nt{First, if $\error^\mmd(\dist)$ is negatively biased.
Second, if plans tend to have more districts $D$ with BVAP just over 50\% ($\bvapmargin_\ddp(\dist)>0$) than just under 50\% ($\bvapmargin_\ddp(\dist)<0$).}
The latter would also help explain the correlation between the number of MMDs and the bias, as maximizing the number of MMDs requires smaller Black majorities on average.

\begin{figure}[h]
    \centering
    \includegraphics[width=\linewidth]{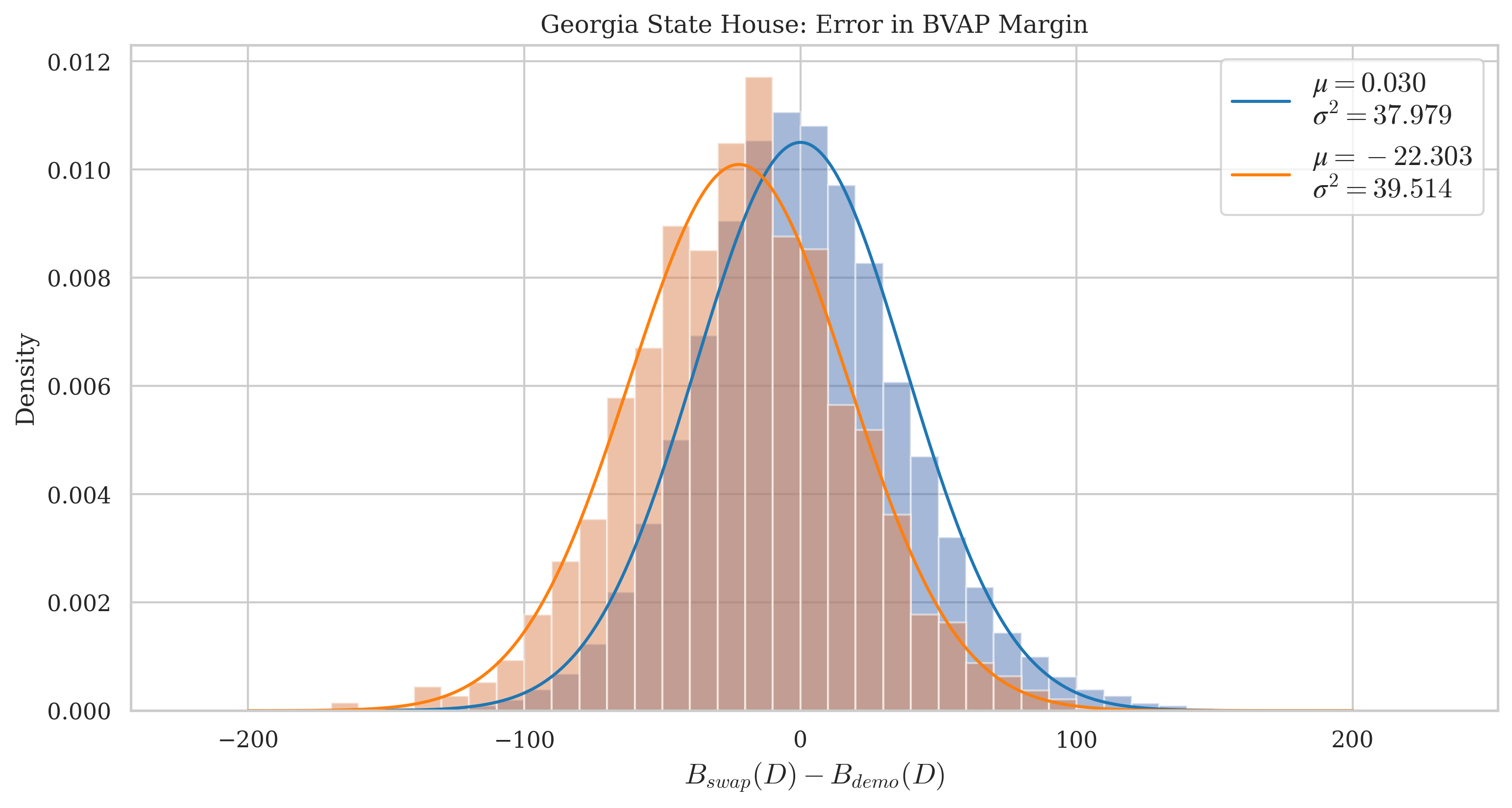}
    \caption{\textbf{Empirical change in BVAP margin.} Change in BVAP margin ($\error^\mmd(\dist) = \bvapmargin_\sfo(\dist) - \bvapmargin_\ddp(\dist)$) among districts in our short bursts ensemble for the Georgia State House of Representatives. The blue histogram includes all districts that were contained within plans in the ensemble. The orange histogram is restricted to districts with Black voting age populations within 196 individuals of half of the total VAP. The threshold of 196 was the maximum error in BVAP margin observed in the ensemble. Note that the noise in BVAP margin for districts that have close to 50\% BVAP is biased in the negative direction.}
    \label{fig:bvap_margin_error}
\end{figure}

Both explanations are borne out in our data for the Georgia state house. 
Figure~\ref{fig:bvap_margin_error} plots $\error^\mmd(\dist)$ for the districts in the short-burst ensemble. 
For districts with BVAP margin close to 0, the mean $\error^\mmd(\dist)$ is $-22.303$ (standard deviation 6.3).\footnote{%
Across all districts, the mean $\error^\mmd(\dist)$ is $0.030$.}

Figure~\ref{fig:GA_black_pop} plots histograms of $\bvapmargin_\ddp(\dist)$ for all districts in both the base and short-burst ensembles.
In the \bursts ensemble, districts with small Black majorities are much more common than districts with small non-Black majorities. This is not true in the \vanilla ensemble which ignores race, helping explain the smaller biases seen in those plans.

\paragraph{A model of MMD discrepancies from BVAP margins}
The discussion above suggests a simple probabilistic model for MMD discrepancies using only BVAP margins, ignoring all other effects (e.g., correlations between errors across districts).
\nt{The model aims to capture the behavior of near-threshold districts whose BVAP is close enough to 50\% that disclosure-avoidance error could change their majority status.}

Given a plan with districts $\dist_i$, sample $\bvapmargin_\sfo(\dist_i) = \bvapmargin_\ddp(\dist_i) + E_i$, where $E_i\sim \normal(\mu,\sigma^2)$ is normally distributed.
\nt{The parameters $\mu$ and $\sigma$ are estimated from the ensemble by looking at districts whose BVAP margins are small enough that a sign flip is possible (i.e., $|\bvapmargin_\ddp(\dist)|\le \max_{\dist'} |\error^\mmd(\dist')|$, the largest observed change in BVAP margin). Intuitively, this conditions on the at-risk districts for which noise can change MMD classification, and avoids letting the many districts with very large margins dominate the fit.}

\begin{figure}
    \centering
    \includegraphics[width=\linewidth]{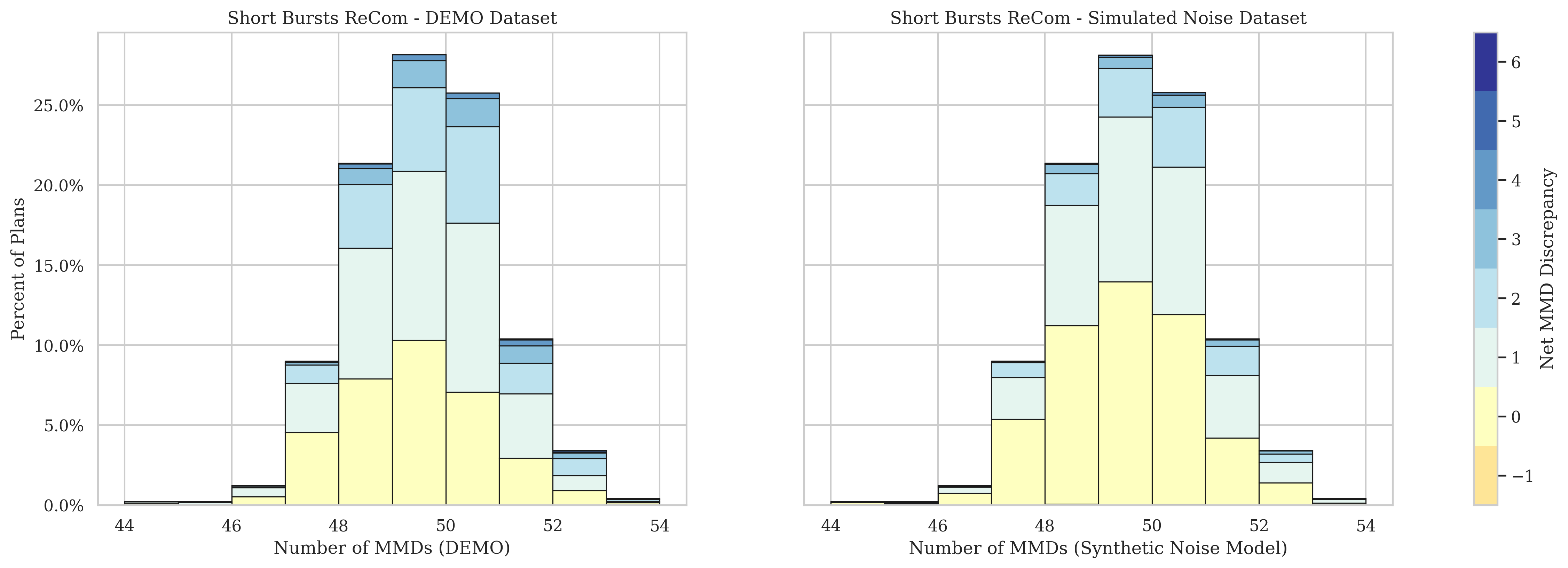}
    \caption{\textbf{Modeling MMD discrepancies in the Georgia state house.} We compare the MMD discrepancies previously observed in Figure~\ref{fig:ga_stacked_hist} to the MMD discrepancies predicted by our synthetic noise model for the Georgia state house.} 
    \label{fig:demo_vs_sim}
\end{figure}

Figure~\ref{fig:demo_vs_sim} plots the results of applying this simulated MMD discrepancy model to the 5,000 Georgia state house plans in our short-bursts ensemble ($\mu =-22.303$, $\sigma^2 = 39.514$), side-by-side with the MMD discrepancies observed when comparing DEMO and SWAP.
The two plots contain exactly the same plans each with the same number of MMDs in DEMO. The only difference is whether MMDs in SWAP are measured against the real data or sampled from the model above.

Qualitatively, the model appears to explain much of the observed 
biases in MMD discrepancies, but not all. 
\nt{The model is deliberately simple and does not capture all features of the observed discrepancies (e.g., non-normality and correlations in $\error^\mmd(\dist)$), so it should be interpreted as a clarifying approximation rather than a full predictive account.}
More work would be needed to quantify the result and to understand what characteristics of the data are driving the remaining bias.

\paragraph{Offsets for Gingles 1 plans}
 As in Section~\ref{sec:popbal}, a natural approach to mitigating MMD discrepancies in the context of Gingles 1 would be to draw illustrative plans whose majority-Black districts have large BVAP margins. Instead of targeting 50\% VAP + 1 person, target 50\% + 1 + \emph{offset}, for some value \emph{offset}$>0$.

We leave a detailed study of the effectiveness of offsets for MMD discrepancies to future work. For now, we remark that this is already common practice. For example, across the eleven illustrative plans used in the recent Supreme Court case \emph{Allen v.\ Milligan (2023)},
the smallest BVAP margin in a majority-Black district was 256 people (50.05\% BVAP)~\cite{smithvmilligan2022sja}, larger than any $\error^\mmd(\dist)$ observed in the Georgia ensemble.

\section{Limitations}
\label{sec:limitations}

\paragraph{Methods}
An important limitation is that we study algorithmically sampled plans, incorporating only a few redistricting criteria (contiguity, population balance, and compactness). 
But real districts are drawn by people as part of a political process, taking account of many factors. One should not read too much into the specific numbers we observe: changing the way we generate districts would likely change the quantitative results.

Some algorithmic methods for sampling plans have the ability to generate representative samples of districting plans under relevant probability distributions for redistricting~\cite{mccartan2023sequential,autry2023metropolized}. For the purposes of this paper, we make no such claims and ignore many of the complexities of practical redistricting. At the same time, from a technical point of view, we believe it is reasonable to offer statistics over a set of interest as a descriptive matter without a need for a distributional claim about the sample. One goal of this paper is to replicate and extend findings from prior work, and to explain the phenomena we observe. For such uses, having a clear, well-defined sample frame may be more useful than strict adherence to human redistricting practice.

Our sample frame does not contain all  permissible plans: it excludes plans that subdivide {precincts}, for instance. 
Nor does it necessarily contain only permissible plans: ignoring political subdivisions and communities of interest in states where required, for example.

It is hard to predict the net effect of these differences on our findings.
As we explain next, breaking up {precincts} would tend to strengthen the effects of the noise. 
On the other hand, the common practice of keeping cities and counties intact would tend to weaken those effects.
Future work could use more sophisticated ensemble analyses to account for individual states' redistricting requirements \citep{becker2021computational, mccartan2022simulated}, or perhaps a dataset of hand-drawn plans.

In order to make the most of our limited computational resources, we chose to use {precincts} as the geounit from which our districts were composed, rather than blocks. This greatly reduced the computational complexity, allowing us to run more chains and study larger geographies. In practice, redistricters may split precincts between neighboring districts (though they often prefer not to).

To get some indication as to how using blocks would have affected our results,
we computed the critical offset and no offset discrepancy rate for three small-enough state geographies using blocks as our base geounit 
(Table \ref{tab:block_chains} in Appendix \ref{app:methods}).
We observe a notable increase in both metrics in all three geographies, indicating that our results may underestimate the impact of the new DAS on redistricting in practice. (On the other hand, keeping cities and counties whole would have the opposite effect.) We hypothesize the difference is greatest where districts are very small, as in the three examined geographies. This aligns with guidance from the Census Bureau suggesting that data users should aggregate block-level data to increase the accuracy of their analyses.

\paragraph{Data}
Comparing the DEMO and SWAP datasets does not isolate the impacts of the 2020 DAS \citep{boyd2022Differential}. It reflects the impact of both the 2010 and 2020 DASes.
This is inherent.
Outside the Census Bureau, it is impossible to directly compare the public 2020 Redistricting Data and the confidential 2020 CEF. 
DEMO was created to aid stakeholders in understanding the effects of the 2020 DAS by comparing it to SWAP.

If we were somehow able to run our analyses using DEMO and the 2010 CEF, nothing in Section~\ref{sec:popbal} would change except possibly in the last paragraph. By using only total population---unaffected by the swapping---we observe the impact of the 2020 DAS alone \cite{Kenny2023Comment}.

On the other hand, Section~\ref{sec:gingles} examines the BVAP as a fraction of the VAP in districts. In 2010, swapping did not affect VAP. But it could affect the BVAP if households with different BVAP were swapped between districts. This is an inherent limitation any work based on comparing the 2010 demonstration and redistricting data \cite{us2021disclosure}.

One way to try to isolate the effect of the 2020 DAS would be to run it ourselves using an alternate dataset (e.g., reconstructed 2010 Census microdata derived from SWAP as in \cite{cohen2022private}). 
But running the 2020 DAS reliably is difficult~\cite{Ballesteros_2025} and beyond the scope of the present work.

In Appendix~\ref{app:swap-swap}, we instead take a step towards isolating the effect of the 2010 DAS, using data provided to us by the authors of~\cite{Ballesteros_2025}. More work would be needed to fully isolate the effects of the two disclosure avoidance systems. Still, based on our analysis in Section~\ref{sec:bvap-margin} and Appendix~\ref{app:swap-swap}, we believe that any differences in the MMD discrepancies arising from the two systems are more likely differences in degree than in kind.

\paragraph{Scope}
Other limitations stem from our scope, rather than our methods. Most obviously, we focused on state legislative redistricting, excluding both Congressional and sub-state redistricting, and on \emph{Gingles 1}, excluding the other parts of the \emph{Gingles} framework and racial-gerrymandering litigation more broadly. So while our results give us hope that discrepancies from disclosure avoidance are tolerable, more serious discrepancies may still lurk elsewhere. 
As a point of comparison, we provide a limited set of results for experiments on Congressional district geographies in 
Table \ref{tab:congress} in Appendix \ref{app:methods}.
Future work could further study Congressional and sub-state redistricting, or extend~\cite{cohen2022private} by quantifying discrepancies for the second and third \emph{Gingles} tests.

Finally, we do not contend with the potential use of the Noisy Measurement Files (NMF) for redistricting (see below). 
While recent work has leveraged the NMF to analyze the error introduced by the 2020 DAS \cite{kenny2024evaluating}, it introduces novel difficulties when applied to redistricting. For example, since hierarchical consistency is not maintained, the population of a district as measured by the NMF can change depending on how its constituent geounits are aggregated. Addressing this issue and evaluating whether the NMF is suitable for redistricting uses is an avenue for future work.

\section{Discussion}
\label{sec:discussion}

In its lawsuit challenging the 2020 DAS, Alabama argued that  One-Person, One-Vote requires balancing populations according to the confidential CEF data, rather than the official Redistricting Data (\emph{Alabama v. Department of Commerce}, 2021). In the face of discrepancies, Alabama concluded that redistricting plans created using the Redistricting Data could be illegal.
\nt{As Alabama put it, this could potentially ``inhibit'' or ``impede'' the redistricting process (\emph{Alabama v.\ Department of Commerce}, 2021).}

What data should redistricting law treat as the ground truth: the data as {collected, edited, and imputed} (the CEF) or as published (the Redistricting Data)?
Our experiments can't answer the question. Instead, they suggest that treating the {CEF} as the ground truth may not greatly affect a redistricter's ability to comply with the law {using only the Redistricting data}. \nt{At the very least, there exist countermeasures for our specific redistricting procedure. To the extent that the finding generalizes, data discrepancies may ``impede'' redistricters by giving them something else to account for, but appears unlikely to ``inhibit'' lawful redistricting altogether.}

At least for state legislative redistricting, it appears that one can use the Redistricting Data to meet OPOV and MMD threshold requirements on the unseen data. Discrepancies do arise, but can be understood and accounted for. For OPOV requirements, using offsets does not make generating state legislative districts more difficult. In particular, all our ensembles in Section~\ref{sec:gingles} used the critical offsets found in Section~\ref{sec:popbal}. And for counting MMDs, the common practice for producing illustrative maps is already to avoid very small margins.

We caution against over-interpreting our results as saying that disclosure avoidance can have no practical impacts on redistricting. Prior work highlights many plausible interactions~\cite{kenny2021use,cohen2022private,wrightirimata}, only a few of which we explore. 
The perturbations from disclosure avoidance (or other sources of error) could conceivably make it harder to challenge a racially gerrymandered plan if, for example, a cracked plan in a racially polarized state appears to have more majority-minority districts in the DEMO data than it has ``effective majority'' districts in reality. Perhaps our probabilistic model of MMD discrepancies might help us reason about this possibility, but we leave that to future work.

Though we are not experts in this area of law, we think it is unlikely that  changes in disclosure avoidance would cause courts to reject the long acknowledged ``legal fiction'' that the Decennial Census exactly reflects the underlying population.
On the other hand, the case for statistical adjustment has only strengthened with the
Census Bureau's recent release of the so-called Noisy Measurement File (NMF)~\cite{nmf}. These are the raw, noisy statistics produced by the 2020 DAS before being post-processed into the required tabular form. By definition, the NMF is statistically more informative than any derived data products, including the Redistricting Data itself. It is conceivable that NMF-based statistical adjustments will find their place redistricting or policymaking. For the moment, this is mere speculation. The official Redistricting Data remains the only basis for redistricting today, to the best of our knowledge.
 
\section*{Acknowledgements}
We thank Peter Rock and Conlan Olson for providing data for our experiments, as well as Micah Altman, John Abowd, and Moon Duchin for valuable feedback. The authors were supported by the DARPA SIEVE program under Agreement No.\ HR00112020021. Preliminary work on this project was supported by the Boston University Center for Antiracist Research.

\bibliographystyle{unsrt}
\bibliography{bib}

\onecolumn
\appendix

\section{Additional Results}
\label{app:methods}

\paragraph{MCMC Convergence tests}
To provide evidence of the convergence of our chains, we report the Gelman-Rubin split-$\hat{R}$ metric and the effective sample size (ESS) for each legislative geography we study in Table~\ref{tab:convergence}. Traditionally, an $\hat{R}$ of below 1.1 has been accepted as evidence of convergence, although more recent work recommends that using a threshold of 1.01 for $\hat{R}$ along with a rank-normalized ESS of at least 400 for more robust results. We test for the convergence of our measurement of population balance (i.e. whether our estimate of the mean of the function $f(\mathcal{P}) = \mathds{1}\left[\mathsf{dev_{swap}}(\mathcal{P}) > 0.1\right]$ has converged, where $\mathcal{P}$ is a plan sampled from the stationary distribution of our Markov chain) and our measurement of MMD discrepancy (i.e. $f(\mathcal{P}) = \mathsf{MMD_{demo}}(\mathcal{P}) - \mathsf{MMD_{swap}}(\mathcal{P})$). For our population balance measurment, we tested convergence using 4 chains of length 1,000,000, subsampling every $10^{th}$ plan to yield an ensemble of 100,000 plans. For the MMD discrepancy measurement, we tested convergence using the 100 ensembles that we generated for each legislative geography as described in Section~\ref{sec:gingles}. We say that a chain shows signs of convergence after a certain number of steps when $\hat{R} < 1.01$ and $\texttt{ESS} > 400$.

All geographies except for the Alabama, Connecticut, West Virginia, and Wyoming state lower houses showed signs of convergence after 1,000,000 steps when measuring population balance. Many geographies did not show signs of converging after 100,000 steps under the MMD discrepancy measurement. However, given our method for creating ensembles with lots of MMDs (sampling maximal plans from each of the 100 chains that we ran), the convergence of the individual chains doesn't have a straightforward relationship to that of the final ensemble. Further work is necessary to characterize the distribution of plans that maximize majority-minority districts.

\begin{table}[]
    \centering
    \scalebox{0.8}{
    \begin{tabular}{lrrrrrrrr}
\toprule
& \multicolumn{4}{c}{State Lower House} & \multicolumn{4}{c}{State Upper House} \\
\cmidrule(lr){2-5}
\cmidrule(lr){6-9}
& \multicolumn{2}{c}{Population Balance} & \multicolumn{2}{c}{MMD} & \multicolumn{2}{c}{Population Balance} & \multicolumn{2}{c}{MMD}\\
\cmidrule(lr){2-3}
\cmidrule(lr){4-5}
\cmidrule(lr){6-7}
\cmidrule(lr){8-9}
    &     $\hat{R}$ &         $\mathsf{ESS}$ &     $\hat{R}$ &        $\mathsf{ESS}$    &     $\hat{R}$ &         $\mathsf{ESS}$ &     $\hat{R}$ &        $\mathsf{ESS}$ \\
\midrule
 AL & \cellcolor{red!20}1.41348 & \cellcolor{red!20}16.2      &          \cellcolor{red!20}1.03605  &      \cellcolor{red!20}1665.2      &   1.00004  & 115308   &   1.01460 &  4993.0   \\
 AK &   1.00053 &   9602.7   & \cellcolor{gray}& \cellcolor{gray}      &   1.00003  &  95286.1 & \cellcolor{gray} & \cellcolor{gray}      \\
 AZ &   1       & 144229      & \cellcolor{gray}& \cellcolor{gray}      &   1.00001  & 135120   & \cellcolor{gray}& \cellcolor{gray}      \\
 AR &   1.00006 &  18532.2    &   1.09077  &   712.0  &   1.00005  &  99684.9 &   1.07575  &    798.1 \\
 CA &   1.00007 &  48475.2    &   1.02305 &  2671.2   &   1        & 131413   & \cellcolor{gray}& \cellcolor{gray}      \\
 CO &   1.00005 &  50360.1    & \cellcolor{gray}& \cellcolor{gray}      &   1.00008  & 118716   & \cellcolor{gray}& \cellcolor{gray}      \\
 CT &   1.04848 &     63.8 & 1.77493 &  149.5     &   1.00001  &  73049.9 &   1.00271   &  30996.1   \\
 DE &   1.00028 &  13237.5    &   1.01678  &   5751.8   &   1.00003  & 115724   &   1.02261  &  2979.9   \\
 FL &   1.00015 &  25729      &   1.03677   &   1722.2  &   1.00001  & 115492   &   1.08056  &  765.7   \\
 GA &   1.00039 &  11449.4    &  1.09305          &         680.1   &   1.00008  &  68435.2 &   1.09921 &  617.7   \\
 HI &   \cellcolor{gray} &    \cellcolor{gray}  &     \cellcolor{gray}       &  \cellcolor{gray}          & \cellcolor{gray}& \cellcolor{gray}   &     \cellcolor{gray}       &   \cellcolor{gray}         \\
 ID &   1.00009 &  58208.3    & \cellcolor{gray}& \cellcolor{gray}      &   1.00005  &  56707.5 & \cellcolor{gray}& \cellcolor{gray}      \\
 IL &   1.00016 &  32520.9    &   1.02456  &   4013.5 &   1.00004  &  73011.1 &   1.00655  &  16931.1    \\
 IN &   1.00029 &  31532.7    &   1.05866  &   1061.5  &   1.00002  &  79920.8 &   1.08445 &   713.4  \\
 IA &   1.00012 &  19643.4    & \cellcolor{gray}& \cellcolor{gray}      &   1.00002  &  56643.6 & \cellcolor{gray}& \cellcolor{gray}      \\
 KS &   1.00022 &  19338      &   1.23501  &   310.5  &   1.00001  &  82320.3 & \cellcolor{gray}& \cellcolor{gray}      \\
 KY &      \cellcolor{red!20}1.00021     &     \cellcolor{red!20}29848.4        &   \cellcolor{red!20}1.02158 &   \cellcolor{red!20}1703.8   &      \cellcolor{red!20}1.00002      &     \cellcolor{red!20}124136.5     &       \cellcolor{gray}     &     \cellcolor{gray}       \\
 LA &   1.00004 &  25932.2    &   1.08096  &    776.0 &   1.00004  &  97275.8 &   1.04140  &   1498.4  \\
 ME &     \cellcolor{red!20}1.00151      &      \cellcolor{red!20}3354.3       &   \cellcolor{gray}         &     \cellcolor{gray}       &   1.00061  &  11535.5 & \cellcolor{gray}& \cellcolor{gray}      \\
 MD &   1.00004 &  40961      &   1.05367  &   1167.7  &   1.00003  &  73312.8 &   1.05580 &  1104.4   \\
 MA &   1.00076 &  10113.5    &   1.04999  &   1335.7  &   1.00009  &  81589.6 &   1.09748   &   648.4  \\
 MI &   1.00008 &  28161.2    &   1.00857  &  13478.6   &   0.999999 & 124932   &   1.02615 &  2555.4   \\
 MN &   1.00034 &  12959.6    & \cellcolor{gray}& \cellcolor{gray}      &   1.00011  &  48035.2 & \cellcolor{gray}& \cellcolor{gray}      \\
 MS &   1.00096 &   7232.6   &   1.09317  &    668.6 &   1.00008  &  53867.2 &   1.04285  &   1481.9  \\
 MO &   1.00011 &  17791.9    &   1.04485   &   1430.1  &   0.999997 & 135122   &   1.14201  &   467.1  \\
 MT &    \cellcolor{red!20}1.00070       &   \cellcolor{red!20}3922.1          &     \cellcolor{gray}       &      \cellcolor{gray}      &     \cellcolor{red!20}1.00005       &       \cellcolor{red!20}29807.9   &  \cellcolor{gray}          &    \cellcolor{gray}        \\
 NE & \cellcolor{black}          & \cellcolor{black}            & \cellcolor{black}           & \cellcolor{black}           &   1.00014  &  53859.5 &   1.31089  &    249.0 \\
 NV &   1.00011 &  64538.9    & \cellcolor{gray}& \cellcolor{gray}      &   1        & 179202   & \cellcolor{gray}& \cellcolor{gray}      \\
 NH &    \cellcolor{gray}       &      \cellcolor{gray}       &       \cellcolor{gray}     &        \cellcolor{gray}    &   1.00969  &    528.9 & \cellcolor{gray}& \cellcolor{gray}      \\
 NJ &   1.00002 & 113039      &   1.02834 &  2232.8   &   1.00004  & 113333   &   1.03377 &   1875.2  \\
 NM &   1.00019 &  25091.2    & \cellcolor{gray}& \cellcolor{gray}      &   1.00005  &  48720.1 & \cellcolor{gray}& \cellcolor{gray}      \\
 NY &   1.00015 &  25506.5    &   1.08593  &    696.3  &   1.00007  &  72752.3 &   1.02001  &  3121.1   \\
 NC &   1.00012 &  20428.8    &   1.04465  &   1463.2  &   1.00008  &  77151.6 &   1.03267  &   1938.9  \\
 ND & \cellcolor{gray} & \cellcolor{gray}      &    \cellcolor{gray}        &     \cellcolor{gray}       &  \cellcolor{gray}&  \cellcolor{gray}  &     \cellcolor{gray}       &     \cellcolor{gray}       \\
 OH &   1       &  40349.8    &   1.04475  &   1415.6  &   0.999999 & 143535   &   1.01950  &   3294.4  \\
 OK &   1.00027 &  22396.2    &   1.12866  &   488.9  &   1.00013  &  66426.5 &   1.00780  &   12631.9  \\
 OR &      \cellcolor{red!20}1.00015     &         \cellcolor{red!20}50394.5    &       \cellcolor{gray}     &    \cellcolor{gray}        &    \cellcolor{red!20}1.00000        &     \cellcolor{red!20}137612.0     &       \cellcolor{gray}     &        \cellcolor{gray}    \\
 PA &   1.00016 &  16996.6    &   1.06915  &   892.7  &   1.00007  &  93319.3 &   1.01360  &   6143.2   \\
 RI &   \cellcolor{red!20}1.00014        &     \cellcolor{red!20}14566.3       & \cellcolor{gray}& \cellcolor{gray}      &        \cellcolor{red!20}1.00006    &       \cellcolor{red!20}58658.8   & \cellcolor{gray}& \cellcolor{gray}      \\
 SC &   1.00032 &  16261.9    &   1.08331  &   769.2  &   1.00001  &  80159.9 &   1.04219  &   1449.7  \\
 SD &   1.00012 &  43806.1    & \cellcolor{gray}& \cellcolor{gray}      &   1.00004  &  50200.9 & \cellcolor{gray}& \cellcolor{gray}      \\
 TN &   1.00011 &  19834.9    &   1.11814  &    543.6 &   1.00001  & 140019   &   1.05512 &  1069.4   \\
 TX &   1.00011 &  23232.5    &   1.01942  &   4100.9   &   1.00003  & 190915   & \cellcolor{gray}& \cellcolor{gray}      \\
 UT &   1.0002  &  29949.4    & \cellcolor{gray}& \cellcolor{gray}      &   1.00001  & 110132   & \cellcolor{gray}& \cellcolor{gray}      \\
 VT &   \cellcolor{gray}        &    \cellcolor{gray}         &   \cellcolor{gray}         &     \cellcolor{gray}       &   1.00001  & 171445   & \cellcolor{gray}& \cellcolor{gray}      \\
 VA &   1.00014 &  23230.8    &   1.04422  &   1397.1  &   1        & 107197   &   1.07987   &    755.4 \\
 WA &   1.00004 &  77180.7    & \cellcolor{gray}& \cellcolor{gray}      &   1.00003  &  73462.3 & \cellcolor{gray}& \cellcolor{gray}      \\
 WV &   1.20887 &     13.0 & \cellcolor{gray}& \cellcolor{gray}      &   1        & 237573   & \cellcolor{gray}& \cellcolor{gray}      \\
 WI &   1.00003 &  29893.9    &   1.02260 &   3215.7  &   1.00004  & 134723   &   1.10688  &   580.7    \\
 WY &   1.08295 &     31.8  &        \cellcolor{gray}    &      \cellcolor{gray}      &   1.00018  &  37894   & \cellcolor{gray}& \cellcolor{gray}      \\
\bottomrule
\end{tabular}
}
\caption{Convergence tests for state legislative district ensembles. Cells that are highlighted in red used block groups as the base unit of geography due, all others used precincts. Cells that are highlighted in gray represent ensembles for which the $\hat{R}$ computation is undefined, either because no plans were found that were balanced in population in our experimental setup, or because every plan within an ensemble contained the same number of MMDs. Cells that are highlighted in black represent legislative geographies that do not exist.}
\label{tab:convergence}
\end{table}

\paragraph{Population Balance and Critical Offsets}
In Table~\ref{tab:offset} we compute no offset discrepancy rates and the critical offsets for state lower and upper house geographies. The critical offsets are computed by repeatedly running chains with a higher offset $\Delta$ from the base acceptable $\tau$. 
For each geography, starting at $\Delta = 0$, we sample an ensemble using a population tolerance threshold in $\textsf{demo}$ of $5\% - \Delta$, and check whether any plan exceeds a 5\% population tolerance. If this there does not exist such a plan in the ensemble, the critical offset is $\Delta$, otherwise we increase $\Delta$ by 0.05\% and repeat the procedure. The average and standard deviation are computed over 10 experimental runs.
We default to $\tau = 5\%$, and $\Delta$ is increased in increments of 0.05\%. We repeat this computation ten times for each geography to observe the variation in critical offset between runs.  The largest deviation we observed between the smallest and largest critical offset for a geography was 0.55\% for the Montana state lower house, with all other geographies having deviations no greater than 0.5\%. We also report what are synthetic model would predict for no offset discrepancy rate and critical offset for each geography, as described in Section~\ref{sec:why-offsets-work}.

\begin{table}[]
    \centering
    \scalebox{0.8}{\begin{tabular}{lrrrrrrrrrr}
\toprule
& \multicolumn{5}{c}{State Lower House} & \multicolumn{5}{c}{State Upper House}\\
\cmidrule(lr){2-6} \cmidrule(lr){7-11}
& \multicolumn{2}{c}{Discrepancy ($\tau=5\%$)} & \multicolumn{3}{c}{Critical Offset ($\Delta$)} & \multicolumn{2}{c}{Discrepancy ($\tau=5\%$)} & \multicolumn{3}{c}{Critical Offset ($\Delta$)}\\
\cmidrule(lr){2-3} \cmidrule(lr){4-6} \cmidrule(lr){7-8} \cmidrule(lr){9-11}
State & Observed & Predicted & Mean $\offset$ & (StDev) & Predicted $\offset$ & Observed & Predicted & Mean $\offset$ & (StDev) & Predicted $\offset$\\
\midrule
 AL      &  \cellcolor{red!20}63.280\% &   \cellcolor{red!20}68.802\% & \cellcolor{red!20}0.535\% & \cellcolor{red!20}(0.045\%) &    \cellcolor{red!20}0.540\% &   19.088\% & 17.617\%   & 0.270\% & (0.024\%) & 0.310\% \\
 AK      &   65.634\% &   70.735\% & 1.580\% & (0.100\%) &    1.670\% &   26.436\% &  29.790\%  & 0.940\% & (0.044\%) & 0.890\% \\
 AZ      &   13.917\% &   13.464\% & 0.235\% & (0.039\%) &    0.265\% &   14.068\% &  13.366\%  & 0.240\% & (0.020\%) & 0.255\% \\
 AR      &   87.448\% &   86.045\% & 0.955\% & (0.057\%) &    1.090\% &   29.745\% &  26.677\%  & 0.435\% & (0.032\%) & 0.460\% \\
 CA      &   23.843\% &   20.163\% & 0.155\% & (0.015\%) &    0.150\% &    8.144\% &  6.425\%  & 0.100\% & (0.000\%) & 0.100\% \\
 CO      &   53.002\% &   51.869\% & 0.550\% & (0.059\%) &    0.600\% &   24.416\% &  22.247\%  & 0.350\% & (0.032\%) & 0.385\% \\
 CT      &  89.200\% &   96.267\% & 0.717\% & (0.062\%) &    1.215\% &   22.958\% &  24.228\%  & 0.370\% & (0.040\%) & 0.400\% \\
 DE      &   58.165\% &   63.885\% & 0.970\% & (0.060\%) &    1.320\% &   24.496\% &  26.760\%  & 0.670\% & (0.033\%) & 0.755\% \\
 FL      &   58.247\% &   53.510\% & 0.350\% & (0.045\%) &    0.350\% &   12.441\% &  10.501\%  & 0.140\% & (0.020\%) & 0.150\% \\
 GA      &   90.185\% &   90.446\% & 0.600\% & (0.050\%) &    0.735\% &   26.967\% &  26.758\%  & 0.275\% & (0.025\%) & 0.310\% \\
 HI      &  \cellcolor{gray} & \cellcolor{gray} &   \cellcolor{gray} &   \cellcolor{gray} & \cellcolor{gray} &  \cellcolor{gray} &   \cellcolor{gray} & \cellcolor{gray} & \cellcolor{gray} & \cellcolor{gray} \\
 ID      &   39.807\% &   40.598\% & 0.725\% & (0.072\%) &    0.785\% &   39.992\% & 40.581\%   & 0.725\% & (0.075\%) & 0.765\% \\
 IL      &   70.243\% &   63.297\% & 0.415\% & (0.039\%) &    0.475\% &   32.335\% &  26.399\%  & 0.260\% & (0.030\%) & 0.270\% \\
 IN      &   67.875\% &   66.219\% & 0.550\% & (0.050\%) &    0.600\% &   30.481\% &  27.136\%  & 0.320\% & (0.033\%) & 0.335\% \\
 IA      &   79.268\% &   77.879\% & 0.765\% & (0.050\%) &    0.865\% &   38.509\% & 35.739\%   & 0.450\% & (0.055\%) & 0.475\% \\
 KS      &   95.661\% &   96.008\% & 1.190\% & (0.070\%) &    1.440\% &   37.253\% & 34.863\%  & 0.555\% & (0.035\%) & 0.565\% \\
 KY      &   \cellcolor{red!20}56.490\% &   \cellcolor{red!20}59.686\% & \cellcolor{red!20}0.480\% & \cellcolor{red!20}(0.046\%) &    \cellcolor{red!20}0.500\% &   \cellcolor{red!20}16.484\% &   \cellcolor{red!20}15.380\% & \cellcolor{red!20}0.260\% & \cellcolor{red!20}(0.037\%) & \cellcolor{red!20}0.230\% \\
 LA      &   76.667\% &   77.106\% & 0.685\% & (0.059\%) &    0.775\% &   24.970\% & 23.534\%   & 0.330\% & (0.033\%) & 0.350\% \\
 ME      & \cellcolor{red!20}95.432\% & \cellcolor{red!20}97.177\% &   \cellcolor{red!20}1.060\% &   \cellcolor{red!20}(0.162\%) & \cellcolor{red!20}1.315\% &   22.543\% & 26.362\% & 0.420\% & (0.075\%) & 0.435\% \\
 MD      &   39.455\% &   40.802\% & 0.380\% & (0.046\%) &    0.420\% &   23.880\% & 24.559\%   & 0.280\% & (0.024\%) & 0.320\% \\
 MA      &   91.418\% &   92.903\% & 0.760\% & (0.070\%) &    0.915\% &   20.662\% &  20.909\%   & 0.275\% & (0.034\%) & 0.315\% \\
 MI      &   51.753\% &   51.264\% & 0.355\% & (0.027\%) &    0.380\% &   11.829\% & 10.398\%   & 0.150\% & (0.000\%) & 0.160\% \\
 MN      &   77.355\% &   81.167\% & 0.655\% & (0.061\%) &    0.690\% &   36.776\% & 39.524\%   & 0.405\% & (0.057\%) & 0.410\% \\
 MS      &   96.175\% &   94.029\% & 1.040\% & (0.058\%) &    1.270\% &   47.674\% & 46.542\%   & 0.600\% & (0.032\%) & 0.640\% \\
 MO      &   92.414\% &   91.548\% & 0.705\% & (0.057\%) &    0.845\% &   17.515\% & 14.543\%   & 0.245\% & (0.015\%) & 0.250\% \\
 MT      &   \cellcolor{red!20}88.118\% &   \cellcolor{red!20}91.207\% & \cellcolor{red!20}1.175\% & \cellcolor{red!20}(0.150\%) &    \cellcolor{red!20}1.355\% &   \cellcolor{red!20}47.250\% &   \cellcolor{red!20}50.706\% & \cellcolor{red!20}0.820\% & \cellcolor{red!20}(0.046\%) & \cellcolor{red!20}0.740\% \\
 NE      & \cellcolor{black} & \cellcolor{black} &   \cellcolor{black} &   \cellcolor{black} & \cellcolor{black} &   51.177\% & 53.399\% & 0.735\% & (0.059\%) & 0.830\% \\
 NV      &   47.556\% &   47.248\% & 0.750\% & (0.045\%) &    0.825\% &   18.552\% &  17.459\% & 0.435\% & (0.039\%) & 0.480\% \\
 NH      & \cellcolor{gray} & \cellcolor{gray} &   \cellcolor{gray} &   \cellcolor{gray} & \cellcolor{gray} &    6.909\% & 11.953\% & 0.240\% & (0.030\%) & 0.285\% \\
 NJ      &   21.662\% &   18.250\% & 0.265\% & (0.039\%) &    0.265\% &   21.368\% & 18.262\%   & 0.255\% & (0.035\%) & 0.270\% \\
 NM      &   81.610\% &   82.258\% & 1.330\% & (0.084\%) &    1.370\% &   52.029\% & 50.167\%   & 0.835\% & (0.045\%) & 0.860\% \\
 NY      &   69.468\% &   63.668\% & 0.345\% & (0.027\%) &    0.380\% &   25.293\% & 20.091\%   & 0.175\% & (0.025\%) & 0.205\% \\
 NC      &   67.835\% &   69.229\% & 0.445\% & (0.027\%) &    0.555\% &   22.654\% & 22.640\%   & 0.260\% & (0.037\%) & 0.285\% \\
 ND      &  \cellcolor{gray} &   \cellcolor{gray} &   \cellcolor{gray} &   \cellcolor{gray} &    \cellcolor{gray} &  \cellcolor{gray} & \cellcolor{gray}   &   \cellcolor{gray} &   \cellcolor{gray} & \cellcolor{gray}\\
 OH      &   52.874\% &   47.655\% & 0.325\% & (0.040\%) &    0.365\% &   12.567\% & 9.318\%   & 0.155\% & (0.015\%) & 0.155\% \\
 OK      &   84.469\% &   84.903\% & 0.970\% & (0.071\%) &    1.050\% &   40.591\% & 40.346\%   & 0.540\% & (0.037\%) & 0.570\% \\
 OR      &   \cellcolor{red!20}41.871\% &   \cellcolor{red!20}43.524\% & \cellcolor{red!20}0.530\% & \cellcolor{red!20}(0.051\%) &    \cellcolor{red!20}0.535\% &   \cellcolor{red!20}15.693\% & \cellcolor{red!20}15.504\%   & \cellcolor{red!20}0.300\% & \cellcolor{red!20}(0.022\%) & \cellcolor{red!20}0.305\% \\
 PA      &   85.595\% &   86.451\% & 0.545\% & (0.057\%) &    0.550\% &   18.750\% & 15.653\%   & 0.170\% & (0.024\%) & 0.195\% \\
 RI      &   \cellcolor{red!20}86.942\% &   \cellcolor{red!20}90.707\% & \cellcolor{red!20}1.795\% & \cellcolor{red!20}(0.117\%) &    \cellcolor{red!20}1.730\% &   \cellcolor{red!20}48.021\% & \cellcolor{red!20}52.211\%   & \cellcolor{red!20}1.225\% & \cellcolor{red!20}(0.093\%) & \cellcolor{red!20}1.040\% \\
 SC      &   84.431\% &   85.243\% & 0.690\% & (0.049\%) &    0.845\% &   29.123\% & 28.846\%   & 0.350\% & (0.032\%) & 0.400\% \\
 SD      &   54.947\% &   54.113\% & 1.020\% & (0.064\%) &    1.110\% &   46.376\% & 50.573\%   & 0.995\% & (0.072\%) & 1.045\% \\
 TN      &   64.002\% &   62.194\% & 0.485\% & (0.071\%) &    0.540\% &   14.440\% & 13.176\%   & 0.225\% & (0.025\%) & 0.230\% \\
 TX      &   62.132\% &   58.322\% & 0.310\% & (0.030\%) &    0.350\% &    8.014\% & 5.564\%   & 0.105\% & (0.015\%) & 0.100\% \\
 UT      &   75.406\% &   74.807\% & 0.880\% & (0.051\%) &    1.005\% &   25.152\% & 23.305\%   & 0.425\% & (0.040\%) & 0.460\% \\
 VT      & \cellcolor{gray} & \cellcolor{gray} &   \cellcolor{gray} &   \cellcolor{gray} & \cellcolor{gray} &    4.730\% & 5.160\% & 0.290\% & (0.020\%) & 0.200\% \\
 VA      &   57.649\% &   57.934\% & 0.405\% & (0.042\%) &    0.485\% &   16.420\% & 15.990\%   & 0.225\% & (0.025\%) & 0.240\% \\
 WA      &   37.025\% &   32.980\% & 0.400\% & (0.039\%) &    0.440\% &   38.125\% & 33.107\%   & 0.400\% & (0.000\%) & 0.445\% \\
 WV      &   66.127\% &   66.311\% & 0.750\% & (0.047\%) &    0.880\% &    8.815\% & 8.598\%   & 0.250\% & (0.032\%) & 0.265\% \\
 WI      &   65.347\% &   64.443\% & 0.555\% & (0.052\%) &    0.570\% &   16.543\% & 13.467\%   & 0.230\% & (0.033\%) & 0.225\% \\
 WY      &  100.000\% &   94.393\% & 2.175\% & (0.144\%) &    2.645\% &   56.729\% & 57.540\%   & 1.440\% & (0.073\%) & 1.485\% \\
\bottomrule
\end{tabular}}
    \caption{\textbf{Empirical and Predicted Discrepancies and Critical Offsets for 93 Legislative Geographies.} 
    Critical offsets are computed as described in Appendix~\ref{app:methods} (averaged over 10 runs). 
    Columns labeled ``Predicted'' are computed by sampling discrepancies (with and without offsets) from the probabilistic model described in Section~\ref{sec:why-offsets-work} using the empirical means and standard deviations of the corresponding  legislative redistricting ensembles.
    Cells that are highlighted in gray represent states where no plans were found that were balanced in population in our experimental setup. Cells that are highlighted in red had precinct data that was incompatible with our experimental setup (either due to large precincts or a lack of complete precinct data) and were computed using block groups as the building blocks of districts instead. Cells that are highlighted in black represent legislative geographies that do not exist. }
    \label{tab:offset}
\end{table}

\paragraph{Robustness to Geounit Type}
All results in the paper were reported for chains using either precincts or census block groups as the base unit of district composition. In Table~\ref{tab:block_chains} we report results for our critical offset computations using chains that are composed from census blocks, the smallest unit of geography defined in the census hierarchy which is sometimes used as the base unit in redistricting. Due to computational constraints, we only report results for three legislative geographies: RI state lower house, RI state upper house, and CT state lower house. In all three geographies, we found that using blocks instead of block groups led to an increase in both no-offset discrepancy rate and critical offset.
See Section~\ref{sec:limitations} for additional discussion about the possible effects of the choice of geounit.

\begin{table}[]
    \centering
    \scalebox{0.8}{
    \begin{tabular}{ccccc}
        \toprule
         & \multicolumn{2}{c}{Block Group/Precinct} & \multicolumn{2}{c}{Block}\\
         \cmidrule(lr){2-3} \cmidrule(lr){4-5}
         & Discrepancy Rate & Critical Offset (\%) & Discrepancy Rate & Critical Offset (\%) \\
        \midrule
        CT SH & 89.20\% & 0.72\% & 98.58\% & $>$2.0\%\\
        RI SH & 86.94\% & 1.80\% & 97.62\% & $>$2.0\%\\
        RI SS & 48.02\% & 1.23\% & 66.14\% & 1.60\% \\
        \bottomrule
    \end{tabular}}
    \caption{Comparison between no offset discrepancy rate and critical offset for three geographies using precincts (Connecticut) or block groups (for Rhode Island legislative geographies) versus blocks as the smallest unit of district construction. We halted the computation of critical offsets at 2.0\%.}
    \label{tab:block_chains}
\end{table}

\paragraph{Robustness to District Type}
We provide metrics from chains run on Congressional districting plans for each state that was apportioned more than one seat in the House of Representatives after the 2010 census. While there is no population deviation that is generally considered de minimis in Congressional redistricting, we adopt the same $\tau = 5\%$ used in our other experiments for the sake of consistency. We find that in every state, the average critical offset over 10 runs, as defined in Section~\ref{sec:popbal}, is less than 0.2\%.

\begin{table}[]
    \centering
    \scalebox{0.8}{
    \begin{tabular}{lrrrr}
\toprule 
& \multicolumn{2}{c}{Population Balance} & \multicolumn{2}{c}{Majority-Minority Districts} \\
\cmidrule(lr){2-3} \cmidrule(lr){4-5}
& \multicolumn{2}{c}{Standard Constraints} & \multicolumn{2}{c}{Short Bursts} \\
\cmidrule(lr){2-3} \cmidrule(lr){4-5} 
 State   & $\Delta$   & DR ($\Delta = 0$)   & Mean Discrep.   & Discrep. Rate\\
\midrule
 AL & 0.10\% &  1.67\% &   0.18 &   17.20\% \\
 AK &  \cellcolor{black} &   \cellcolor{black} & \cellcolor{black}    &     \cellcolor{black} \\
 AZ & 0.10\% &  2.02\% &  \cellcolor{gray} & \cellcolor{gray} \\
 AR & 0.10\% &  0.74\% &  \cellcolor{gray} & \cellcolor{gray} \\
 CA & 0.13\% & 12.81\% &  \cellcolor{gray} & \cellcolor{gray} \\
 CO & 0.11\% &  1.84\% &  \cellcolor{gray} & \cellcolor{gray} \\
 CT & 0.10\% &  0.77\% &  \cellcolor{gray} & \cellcolor{gray} \\
 DE &  \cellcolor{black} &   \cellcolor{black} & \cellcolor{black}    &     \cellcolor{black} \\
 FL & 0.12\% &  6.79\% &   0.10 &   10.10\% \\
 GA & 0.10\% &  3.23\% &   0.47 &   44.40\% \\
 HI & 0.10\% &  0.26\% &  \cellcolor{gray} & \cellcolor{gray} \\
 ID & 0.10\% &  0.18\% &  \cellcolor{gray} & \cellcolor{gray} \\
 IL & 0.12\% &  5.93\% &   0.15 &   13.70\% \\
 IN & 0.10\% &  2.12\% &  \cellcolor{gray} & \cellcolor{gray} \\
 IA & 0.10\% &  0.53\% &  \cellcolor{gray} & \cellcolor{gray} \\
 KS & 0.10\% &  1.00\% &  \cellcolor{gray} & \cellcolor{gray} \\
 KY & \cellcolor{red!20}0.10\% &  \cellcolor{red!20}0.88\% &  \cellcolor{gray} & \cellcolor{gray} \\
 LA & 0.10\% &  1.18\% &   0.15 &   15.30\% \\
 ME & 0.10\% &  0.01\% &  \cellcolor{gray} & \cellcolor{gray} \\
 MD & 0.10\% &  1.38\% &   0.19 &   18.40\% \\
 MA & 0.10\% &  1.78\% &  \cellcolor{gray} & \cellcolor{gray} \\
 MI & 0.10\% &  2.48\% &   0.32 &   27.90\% \\
 MN & 0.10\% &  1.48\% &  \cellcolor{gray} & \cellcolor{gray} \\
 MS & 0.10\% &  0.85\% &   0.01 &    0.90\% \\
 MO & 0.10\% &  1.67\% &  \cellcolor{gray} & \cellcolor{gray} \\
 MT &  \cellcolor{black} &   \cellcolor{black} & \cellcolor{black}    &     \cellcolor{black} \\
 NE & 0.10\% &  0.46\% & \cellcolor{black}    &     \cellcolor{black} \\
 NV & 0.12\% &  1.04\% &  \cellcolor{gray} & \cellcolor{gray} \\
 NH & 0.10\% &  0.04\% &  \cellcolor{gray} & \cellcolor{gray} \\
 NJ & 0.11\% &  3.51\% &   0.17 &   16.60\% \\
 NM & 0.10\% &  0.64\% &  \cellcolor{gray} & \cellcolor{gray} \\
 NY & 0.11\% &  6.80\% &   0.23 &   21.80\% \\
 NC & 0.10\% &  3.25\% &   0.05 &    5.30\% \\
 ND &  \cellcolor{black} &   \cellcolor{black} & \cellcolor{black}    &     \cellcolor{black} \\
 OH & 0.10\% &  3.95\% &   0.02 &    2.40\% \\
 OK & 0.10\% &  1.22\% &  \cellcolor{gray} & \cellcolor{gray} \\
 OR & \cellcolor{red!20}0.10\% &  \cellcolor{red!20}0.78\% &  \cellcolor{gray} & \cellcolor{gray} \\
 PA & 0.10\% &  4.28\% &   0.25 &   25.00\% \\
 RI & \cellcolor{red!20}0.10\% &  \cellcolor{red!20}0.31\% &  \cellcolor{gray} & \cellcolor{gray} \\
 SC & 0.10\% &  1.54\% &   0.26 &   25.70\% \\
 SD &  \cellcolor{black} &   \cellcolor{black} & \cellcolor{black}    &     \cellcolor{black} \\
 TN & 0.10\% &  1.98\% &   0.14 &   14.30\% \\
 TX & 0.14\% &  9.08\% &   0.01 &    1.40\% \\
 UT & 0.13\% &  0.86\% &  \cellcolor{gray} & \cellcolor{gray} \\
 VT &  \cellcolor{black} &   \cellcolor{black} & \cellcolor{black}    &     \cellcolor{black} \\
 VA & 0.10\% &  2.14\% &   0.28 &   27.50\% \\
 WA & 0.19\% &  3.16\% &  \cellcolor{gray} & \cellcolor{gray} \\
 WV & 0.10\% &  0.42\% &  \cellcolor{gray} & \cellcolor{gray} \\
 WI & 0.10\% &  1.80\% &  \cellcolor{gray} & \cellcolor{gray} \\
 WY &  \cellcolor{black} &   \cellcolor{black} & \cellcolor{black}    &     \cellcolor{black} \\
\bottomrule
\end{tabular}
}

    \caption{Metrics for ensembles of Congressional district plans. Cells highlighted in black represent states that were apportioned a single congressional representative after the 2010 census, cells highlighted in gray represent geographies for which no majority-Black districts were found by our chains, and cells highlighted in red represent chains that were run using block groups as the base unit of geography due to missing precinct data. For each ensemble sampled with standard constraints, we calculate the critical offset ($\Delta$) and the no-offset discrepancy rate (i.e. the proportion of plans exceeding a 5\% population balance threshold under $\mathsf{SWAP}$). Critical offsets are averaged over 10 runs. For each ensemble sampled using short burst optimization, we calculate the mean MMD discrepancy and the discrepancy rate (i.e. the proportion of plans with a non-zero MMD discrepancy). }
    \label{tab:congress}
\end{table}

\paragraph{Robustness to Minority Group}
We compare our results from Section 5 to state lower house chains optimized to increase the number of majority-Hispanic districts in a given plan, rather than majority-Black districts. Each optimized ensemble is constructed by running 100 ReCom chains with short bursts optimization to generate plans with many majority Hispanic districts, and sampling 10 plans from each chain with the highest number of majority Hispanic districts to be included in the ensemble. Each chain uses a population tolerance threshold of $5\% - \offset$, where $\offset$ is the  critical offset for that legislative geography, as computed in Section~\ref{sec:popbal}. The magnitudes of discrepancies we observed for majority-Hispanic districts are similar in magnitude to those of majority-Black districts. For example, the largest mean discrepancy we observed for majority-Hispanic districts was 1.093 in the New Mexico State Lower House, while the larges we observed for majority-Black districts was 1.431 in the Mississippi State Lower House.

\begin{table}[]
    \centering
    \scalebox{0.8}{
    \begin{tabular}{lrrrr}
\toprule
& \multicolumn{2}{c}{State Lower House} & \multicolumn{2}{c}{State Upper House} \\
\cmidrule(lr){2-3} \cmidrule(lr){4-5}
 State   & Non-Zero Discrep. Rate   & Mean Discrep.    & Non-Zero Discrep. Rate   & Mean Discrep.    \\
\midrule
 AL      & \cellcolor{gray}         & \cellcolor{gray} & \cellcolor{gray}         & \cellcolor{gray} \\
 AK      & \cellcolor{gray}         & \cellcolor{gray} & \cellcolor{gray}         & \cellcolor{gray} \\
 AZ      & 32.5\%                    & 0.387            & 27.4\%                    & 0.336            \\
 AR      & \cellcolor{gray}         & \cellcolor{gray} & \cellcolor{gray}         & \cellcolor{gray} \\
 CA      & 47.4\%                    & 0.601            & 40.1\%                    & 0.523            \\
 CO      & 58.0\%                    & 0.717            & 22.2\%                    & 0.234            \\
 CT      & 1.1\%                     & 0.011            & 0.0\%                     & 0.000            \\
 DE      & \cellcolor{gray}         & \cellcolor{gray} & \cellcolor{gray}         & \cellcolor{gray} \\
 FL      & 27.1\%                    & 0.286            & 6.5\%                     & 0.066            \\
 GA      & 0.3\%                     & 0.003            & \cellcolor{gray}         & \cellcolor{gray} \\
 HI      & \cellcolor{gray}         & \cellcolor{gray} & \cellcolor{gray}         & \cellcolor{gray} \\
 ID      & \cellcolor{gray}         & \cellcolor{gray} & \cellcolor{gray}         & \cellcolor{gray} \\
 IL      & 32.2\%                    & 0.344            & 3.4\%                     & 0.034            \\
 IN      & \cellcolor{gray}         & \cellcolor{gray} & \cellcolor{gray}         & \cellcolor{gray} \\
 IA      & \cellcolor{gray}         & \cellcolor{gray} & \cellcolor{gray}         & \cellcolor{gray} \\
 KS      & 7.5\%                     & 0.076            & 5.2\%                     & 0.052            \\
 KY      & \cellcolor{gray}         & \cellcolor{gray} & \cellcolor{gray}         & \cellcolor{gray} \\
 LA      & \cellcolor{gray}         & \cellcolor{gray} & \cellcolor{gray}         & \cellcolor{gray} \\
 ME      & \cellcolor{gray}         & \cellcolor{gray} & \cellcolor{gray}         & \cellcolor{gray} \\
 MD      & 0.8\%                     & 0.008            & 0.0\%                     & 0.000            \\
 MA      & 29.2\%                    & 0.312            & \cellcolor{gray}         & \cellcolor{gray} \\
 MI      & \cellcolor{gray}         & \cellcolor{gray} & \cellcolor{gray}         & \cellcolor{gray} \\
 MN      & \cellcolor{gray}         & \cellcolor{gray} & \cellcolor{gray}         & \cellcolor{gray} \\
 MS      & \cellcolor{gray}         & \cellcolor{gray} & \cellcolor{gray}         & \cellcolor{gray} \\
 MO      & \cellcolor{gray}         & \cellcolor{gray} & \cellcolor{gray}         & \cellcolor{gray} \\
 MT      & \cellcolor{gray}         & \cellcolor{gray} & \cellcolor{gray}         & \cellcolor{gray} \\
 NE      & \cellcolor{black}         & \cellcolor{black} & 0.0\%                     & 0.000            \\
 NV      & 35.1\%                    & 0.430            & 55.9\%                    & 0.664            \\
 NH      & \cellcolor{gray}         & \cellcolor{gray} & \cellcolor{gray}         & \cellcolor{gray} \\
 NJ      & 20.6\%                    & 0.226            & 19.1\%                    & 0.206            \\
 NM      & 67.3\%                    & 1.093            & 56.4\%                    & 0.811            \\
 NY      & 26.7\%                    & 0.322            & 22.2\%                    & 0.238            \\
 NC      & \cellcolor{gray}         & \cellcolor{gray} & \cellcolor{gray}         & \cellcolor{gray} \\
 ND      & \cellcolor{gray}         & \cellcolor{gray} & \cellcolor{gray}         & \cellcolor{gray} \\
 OH      & \cellcolor{gray}         & \cellcolor{gray} & \cellcolor{gray}         & \cellcolor{gray} \\
 OK      & 10.3\%                    & 0.087            & 7.9\%                     & -0.079           \\
 OR      & \cellcolor{gray}         & \cellcolor{gray} & \cellcolor{gray}         & \cellcolor{gray} \\
 PA      & 23.7\%                    & 0.253            & \cellcolor{gray}         & \cellcolor{gray} \\
 RI      & \cellcolor{gray}         & \cellcolor{gray} & \cellcolor{gray}         & \cellcolor{gray} \\
 SC      & \cellcolor{gray}         & \cellcolor{gray} & \cellcolor{gray}         & \cellcolor{gray} \\
 SD      & \cellcolor{gray}         & \cellcolor{gray} & \cellcolor{gray}         & \cellcolor{gray} \\
 TN      & \cellcolor{gray}         & \cellcolor{gray} & \cellcolor{gray}         & \cellcolor{gray} \\
 TX      & 57.2\%                    & 0.743            & 11.4\%                    & 0.118            \\
 UT      & \cellcolor{gray}         & \cellcolor{gray} & \cellcolor{gray}         & \cellcolor{gray} \\
 VT      & \cellcolor{gray}         & \cellcolor{gray} & \cellcolor{gray}         & \cellcolor{gray} \\
 VA      & \cellcolor{gray}         & \cellcolor{gray} & \cellcolor{gray}         & \cellcolor{gray} \\
 WA      & 60.8\%                    & 0.719            & 55.7\%                    & 0.648            \\
 WV      & \cellcolor{gray}         & \cellcolor{gray} & \cellcolor{gray}         & \cellcolor{gray} \\
 WI      & 16.9\%                    & 0.172            & \cellcolor{gray}         & \cellcolor{gray} \\
 WY      & \cellcolor{gray}         & \cellcolor{gray} & \cellcolor{gray}         & \cellcolor{gray} \\
\bottomrule
\end{tabular}
}

    \caption{Metrics for ensembles optimized to increase the number of majority-Hispanic districts. We show the mean majority-Hispanic district discrepancy and the discrepancy rate (i.e. the proportion of plans with a non-zero MMD discrepancy). Gray cells represent geographies for which our ensembles did not contain any majority-Hispanic districts in either DEMO or SWAP. Black cells represent legislative geographies that do not exist.}
    \label{tab:hisp}
\end{table}

\paragraph{MMD Discrepancies Across Geographies}
Table~\ref{tab:mmd_discreps} contains the mean MMD discrepancy and non-zero discrepancy rate for optimized ensembles across 52 legislative geographies. Each optimized ensemble is constructed by running 100 ReCom chains with short bursts optimization to generate plans with many majority-Black districts, and sampling 10 plans from each chain with the highest number of majority-Black districts to be included in the ensemble. Each chain uses a population tolerance threshold of $5\% - \offset$, where $\offset$ is the  critical offset for that legislative geography, as computed in Section~\ref{sec:popbal}. The legislative geographies included are those for which our chains found at least one plan with at least one majority-minority district.

\begin{table}[]
    \centering
    \scalebox{0.8}{\begin{tabular}{lrrrr}
\toprule
& \multicolumn{4}{c}{Short Bursts} \\
\cmidrule(lr){2-5} 
 State   & Non-Zero Discrep.\ Rate   & Mean Discrep.\    & Non-Zero Discrep.\ Rate   & Mean Discrep.\    \\
\midrule
 AL      & 57.5\%                    & 0.822            & 25.3\%                    & 0.293            \\
 AK      & \cellcolor{gray}         & \cellcolor{gray} & \cellcolor{gray}         & \cellcolor{gray} \\
 AZ      & \cellcolor{gray}         & \cellcolor{gray} & \cellcolor{gray}         & \cellcolor{gray} \\
 AR      & 37.2\%                    & 0.470            & 54.2\%                    & 0.715            \\
 CA      & 0.9\%                     & 0.009            & \cellcolor{gray}         & \cellcolor{gray} \\
 CO      & \cellcolor{gray}         & \cellcolor{gray} & \cellcolor{gray}         & \cellcolor{gray} \\
 CT      & 11.6\%         & 0.126 & 6.1\%                     & 0.061            \\
 DE      & 11.7\%                    & 0.120            & 4.3\%                     & 0.045            \\
 FL      & 23.4\%                    & 0.256            & 7.6\%                     & 0.076            \\
 GA      & 65.6\%         & 1.027 & 34.7\%                    & 0.433            \\
 HI      & \cellcolor{gray}         & \cellcolor{gray} & \cellcolor{gray}         & \cellcolor{gray} \\
 ID      & \cellcolor{gray}         & \cellcolor{gray} & \cellcolor{gray}         & \cellcolor{gray} \\
 IL      & 23.3\%                    & 0.270            & 8.9\%                     & 0.095            \\
 IN      & 31.5\%                    & 0.364            & 28.2\%                    & 0.291            \\
 IA      & \cellcolor{gray}         & \cellcolor{gray} & \cellcolor{gray}         & \cellcolor{gray} \\
 KS      & 39.2\%                    & 0.415            & \cellcolor{gray}         & \cellcolor{gray} \\
 KY      & \cellcolor{red!20}2.4\%         & \cellcolor{red!20}0.024 & \cellcolor{gray}         & \cellcolor{gray} \\
 LA      & 68.3\%                    & 1.122            & 43.7\%                    & 0.553            \\
 ME      & \cellcolor{gray}         & \cellcolor{gray} & \cellcolor{gray}         & \cellcolor{gray} \\
 MD      & 17.6\%                    & 0.191            & 16.6\%                    & 0.181            \\
 MA      & 17.6\%                    & 0.181            & 5.4\%                     & 0.054            \\
 MI      & 11.3\%                    & 0.120            & 6.4\%                     & 0.067            \\
 MN      & \cellcolor{gray}         & \cellcolor{gray} & \cellcolor{gray}         & \cellcolor{gray} \\
 MS      & 78.9\%                    & 1.431            & 53.7\%                    & 0.747            \\
 MO      & 32.9\%                    & 0.405            & 58.3\%                    & 0.790            \\
 MT      & \cellcolor{gray}         & \cellcolor{gray} & \cellcolor{gray}         & \cellcolor{gray} \\
 NE      & \cellcolor{black}         & \cellcolor{black} & 38.9\%                    & 0.389            \\
 NV      & \cellcolor{gray}         & \cellcolor{gray} & \cellcolor{gray}         & \cellcolor{gray} \\
 NH      & \cellcolor{gray}         & \cellcolor{gray} & \cellcolor{gray}         & \cellcolor{gray} \\
 NJ      & 11.0\%                    & 0.114            & 10.1\%                    & 0.107            \\
 NM      & \cellcolor{gray}         & \cellcolor{gray} & \cellcolor{gray}         & \cellcolor{gray} \\
 NY      & 42.7\%                    & 0.533            & 17.5\%                    & 0.187            \\
 NC      & 49.2\%                    & 0.649            & 27.8\%                    & 0.315            \\
 ND      & \cellcolor{gray}         & \cellcolor{gray} & \cellcolor{gray}         & \cellcolor{gray} \\
 OH      & 24.8\%                    & 0.289            & 24.3\%                    & 0.261            \\
 OK      & 34.7\%                    & 0.404            & 15.0\%                    & 0.150            \\
 OR      & \cellcolor{gray}         & \cellcolor{gray} & \cellcolor{gray}         & \cellcolor{gray} \\
 PA      & 20.2\%                    & 0.220            & 14.3\%                    & 0.149            \\
 RI      & \cellcolor{gray}         & \cellcolor{gray} & \cellcolor{gray}         & \cellcolor{gray} \\
 SC      & 71.3\%                    & 1.203            & 59.3\%                    & 0.847            \\
 SD      & \cellcolor{gray}         & \cellcolor{gray} & \cellcolor{gray}         & \cellcolor{gray} \\
 TN      & 30.5\%                    & 0.325            & 26.0\%                    & 0.291            \\
 TX      & 9.2\%                     & 0.096            & \cellcolor{gray}         & \cellcolor{gray} \\
 UT      & \cellcolor{gray}         & \cellcolor{gray} & \cellcolor{gray}         & \cellcolor{gray} \\
 VT      & \cellcolor{gray}         & \cellcolor{gray} & \cellcolor{gray}         & \cellcolor{gray} \\
 VA      & 31.6\%                    & 0.367            & 53.1\%                    & 0.724            \\
 WA      & \cellcolor{gray}         & \cellcolor{gray} & \cellcolor{gray}         & \cellcolor{gray} \\
 WV      & \cellcolor{gray}         & \cellcolor{gray} & \cellcolor{gray}         & \cellcolor{gray} \\
 WI      & 20.3\%                    & 0.218            & 46.3\%                    & 0.493            \\
 WY      & \cellcolor{gray}         & \cellcolor{gray} & \cellcolor{gray}         & \cellcolor{gray} \\
\bottomrule
\end{tabular}}
    \caption{Mean majority-Black district discrepancy and non-zero discrepancy rate for short-burst optimized ensembles across 52 state legislative geographies. Gray cells represent geographies for which our ensembles did not contain any majority-Black districts in either DEMO or SWAP. Black cells represent legislative geographies that do not exist. Red cells represent chains that were run using block groups as the base unit of geography due to missing precinct data.}
    \label{tab:mmd_discreps}
\end{table}

\begin{figure}[h!]
    \centering
    \includegraphics[width=\linewidth]{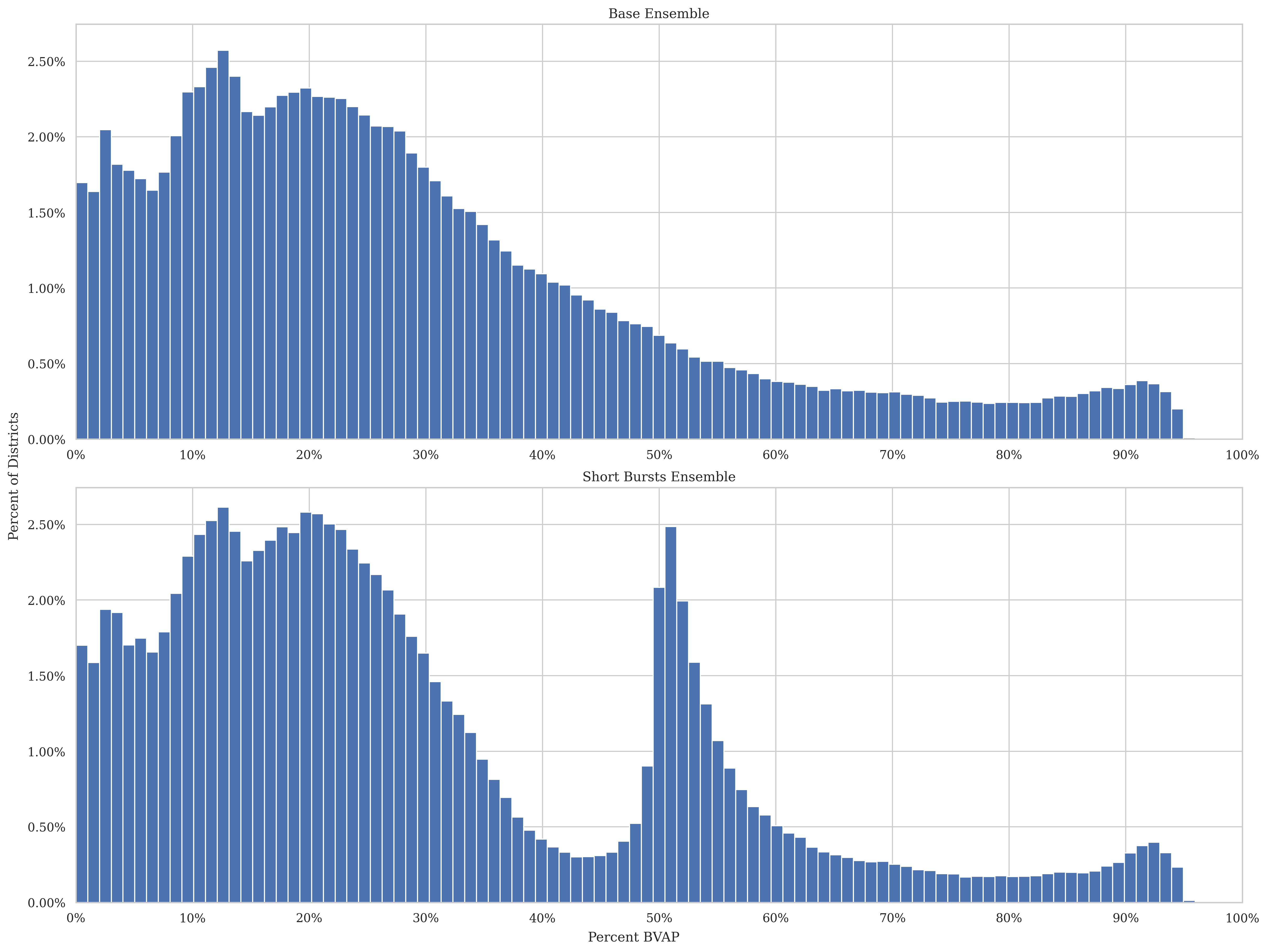}
    \caption{\textbf{Histograms of \% BVAP population for districts in two Georgia state house ensembles.} The data include all distinct districts in our base (top) and short bursts (bottom) ensembles.
    The short bursts ensemble, which is designed to produce plans with many MMDs, samples districts with small Black majorities much more often than the base ensemble which ignores race.}
    \label{fig:GA_black_pop}
\end{figure}

\paragraph{Distribution of BVAP Population in Standard and Optimized Ensembles.}
Figure~\ref{fig:GA_black_pop} plots histograms of the \%BVAP for all plans sampled in the base ensemble and the short bursts ensemble. The short burst optimization technique samples heavily from plans that are just over 50\% BVAP.

\section{Towards Disentangling the Impact of the 2010 and 2020 Disclosure Avoidance Systems}
\label{app:swap-swap}

We would like to understand the impact of the 2020 DAS on MMD discrepancies. 
Unfortunately, comparing DEMO and SWAP reflects the combined effects of the 2010 DAS and the 2020 DAS. 
This is because DEMO is the result of applying the 2020 DAS to the confidential 2010 CEF data, while SWAP is the result of applying the 2010 DAS to the confidential 2010 CEF data.

One way to try to isolate the effect of the 2020 DAS would be to run it ourselves using reconstructed 2010 Census microdata derived from SWAP (as in \cite{cohen2022private}). 
But running the 2020 DAS reliably is difficult~\cite{Ballesteros_2025} and beyond the scope of the present work.

Instead, we try to isolate the effect of the 2010 DAS using a dataset provided to us by the authors of \cite{Ballesteros_2025} (personal communication, September 2025). 
The dataset is the result of applying an implementation of \emph{swapping} (designed to approximate the 2010 DAS as closely as possible using only publicly-available information) to reconstructed 2010 Census microdata for Georgia, as described in \cite{Ballesteros_2025}.
Hence, this dataset---which we call \emph{SWAP-SWAP}---is the result of applying the swapping-based 2010 DAS to the 2010 CEF, then reconstructing it and applying an independent implementation of swapping.

\begin{figure}
    \centering
    \includegraphics[width=\linewidth]{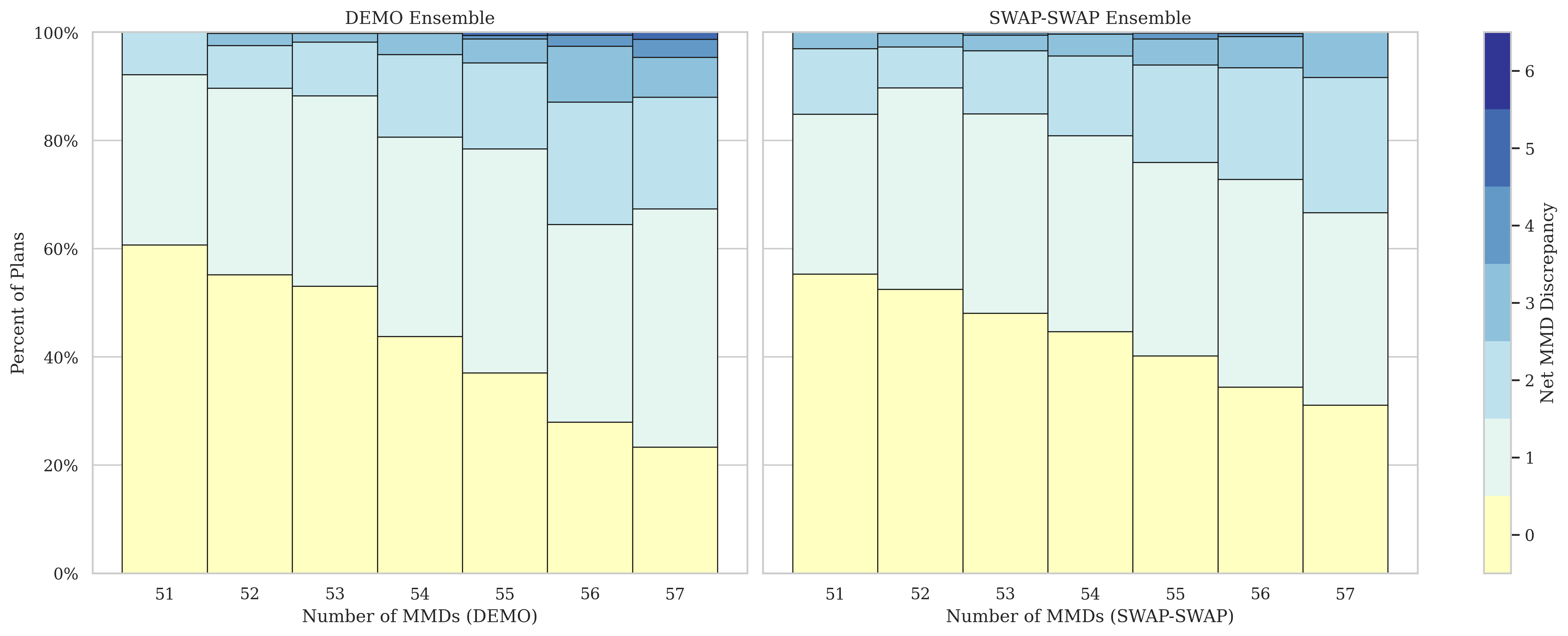}
    \caption{\textbf{MMD discrepancies in SWAP-SWAP.} Distribution of MMD discrepancies (relative to SWAP) for Georgia state house using the DEMO and SWAP-SWAP datasets. The number of MMDs on the horizontal axis is measured using the dataset used to generate the ensembles (DEMO or SWAP-SWAP), and the percent of plans normalized to 100\% for each bin is displayed on the vertical axis. We generate ensembles using the method described in Section~\ref{sec:shortburst}, running 100 chains for each dataset.}
    \label{fig:demo_vs_swap}
\end{figure}

We generate a new ensembles of plans using the SWAP-SWAP and DEMO datasets (using short-bursts optimization for \recom with 100 chains in parallel). For each ensemble, we measure the MMD discrepancies relative to SWAP. 
(For this analysis only, we count majority-Black districts using the total population rather than the voting age population, because voting age data was not included in the SWAP-SWAP dataset.)
The results are plotted in Figure~\ref{fig:demo_vs_swap}. The similarities are clear. In both ensembles, the number of MMDs according the dataset used to generate the plans (DEMO, SWAP-SWAP) tends to be greater than the number of MMDs according to the unseen dataset (SWAP). Moreover, the bias grows with the number of MMDs.

\nt{
Going a step further, SWAP-SWAP population data makes it possible to estimate what Black population margins might look like in the CEF. 
This holds under a strong assumption on the fidelity of the SWAP-SWAP data.\footnote{It appears difficult or impossible to validate this assumption for the same reasons that measuring the effects the 2010 swapping-based DAS is difficult: the swapping algorithm and the CEF input data is secret, known only to the Census Bureau. } Namely, the analysis below requires the differences between tabulations in SWAP-SWAP and SWAP be distributed similarly to the differences between tabulations in SWAP and the 2010 CEF. 
Adapting the notation introduced in Section~\ref{sec:bvap-margin}, 
\begin{align}
    B_\mathsf{swap\text{-}swap}(D) - B_\mathsf{swap}(D) \approx B_{\mathsf{swap}}(D) - B_\mathsf{cef}(D),
\end{align}
where $B_{\mathsf{data}}(D)$ is the Black population margin (i.e., $\pop^\mathsf{Black}_\mathsf{data}(D) - 0.5\cdot\pop^\mathsf{Black}_\mathsf{dara}(D)$) in district $D$ as measured in dataset DATA. This differs slightly from the definition of $B_\mathsf{data}(D)$ in Section~\ref{sec:bvap-margin}, which considers voting age population rather than total population. This is due to the lack of voting age information in SWAP-SWAP, as mentioned above.}

\begin{figure}[h]
    \centering
    \includegraphics[width=\linewidth]{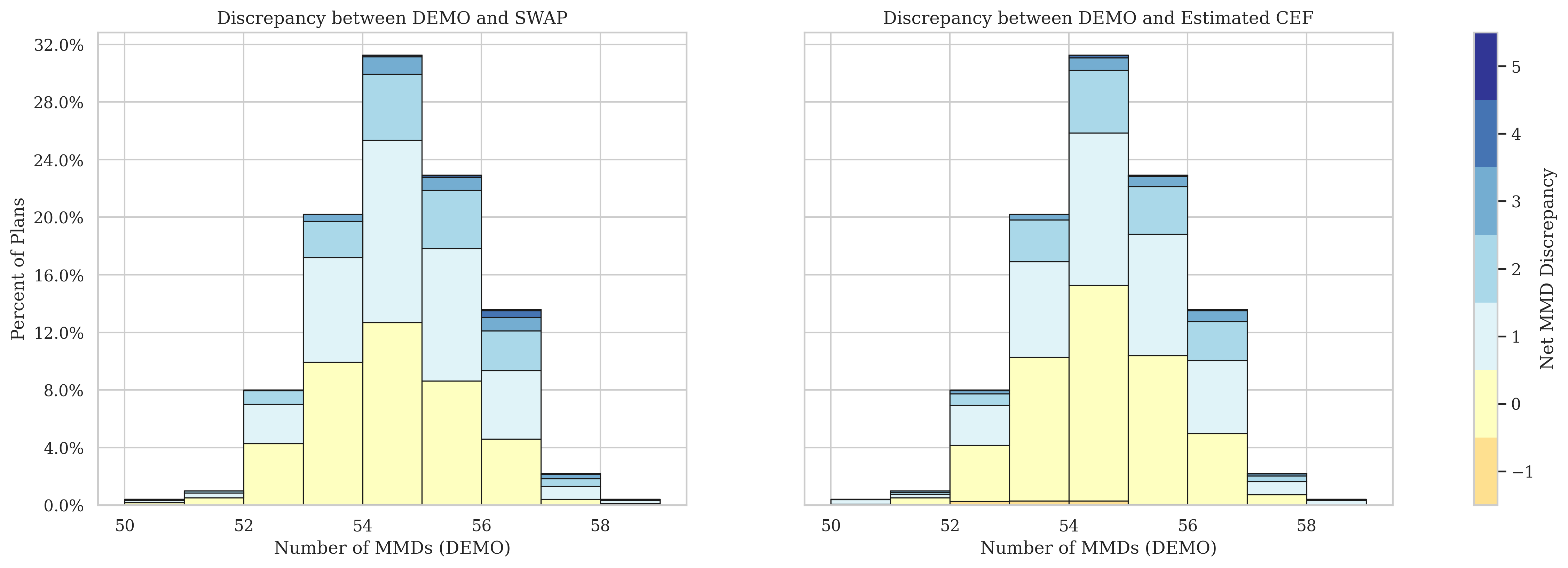}
    \caption{\textbf{MMD discrepancies in Reconstructed CEF Data.} Distribution of MMD discrepancies for an ensemble using short bursts optimization to sample plans with many MMDs in DEMO. On the left, we display the MMD discrepancy between DEMO and SWAP. On the right, we display the MMD discrepancy between DEMO and our estimation of the 2010 CEF, as described in Appendix~\ref{app:swap-swap}.}
    \label{fig:cef_reconstruction}
\end{figure}

\nt{Assuming the above, we can approximate Black population margins in the 2010 CEF directly from the margins we measured using SWAP and SWAP-SWAP:
\begin{align}
    B_\mathsf{cef}(D) &= B_\mathsf{swap}(D) + \bigl(B_\mathsf{cef}(D) - B_\mathsf{swap}(D)\bigr)\\
    &\approx B_\mathsf{swap}(D) + \bigl(B_\mathsf{swap}(D) - B_\mathsf{swap\text{-}swap}(D)\bigr)\\ 
    &= 2B_\mathsf{swap}(D) - B_\mathsf{swap\text{-}swap}(D)
\end{align}
Using these estimates, we plot MMD discrepancies between DEMO and the above estimate of CEF in Figure~\ref{fig:cef_reconstruction}. Discrepancies between DEMO and CEF are qualitatively similar to discrepancies between DEMO and SWAP, although slightly smaller in magnitude. This result suggests that our observations in Section~\ref{sec:gingles} would have been similar had we been using the 2010 CEF rather than SWAP.

To reiterate, the analysis above assumes that the implementation of swapping used to create our SWAP-SWAP dataset closely approximates the 2010 DAS.
}
More work would be needed to fully disentangle the effects of the 2020 and 2010 disclosure avoidance systems, even setting aside the extent to which comparing SWAP-SWAP to SWAP is a faithful representation of the effects of the 2010 DAS.
For example, it is not clear whether the similarities between the to plots in Figure~\ref{fig:demo_vs_swap} mean that the effect of the 2020 DAS is smaller, larger, or similar to the effects of swapping. 
Instead, we remark only that any differences in the MMD discrepancies arising from the two systems appear more likely differences in degree than in kind.

\end{document}